\documentclass[aps,prl,twocolumn,preprintnumbers,superscriptaddress,nofootinbib]{revtex4}
\usepackage[utf8]{inputenc}
\usepackage{graphicx}
\usepackage{psfrag}
\usepackage{color}
\usepackage{amssymb}
\usepackage{amsmath}
\usepackage{braket}
\usepackage{epstopdf}
\begin{document}
\title{Tomography of light mesons in light-cone quark model}
\author{Satvir Kaur}
\email{satvirkaur578@gmail.com}
\affiliation{Department of Physics, Dr. B. R. Ambedkar National Institute of Technology, Jalandhar 144011, India}

\author{Narinder Kumar}
\email{narinderhep@gmail.com}
\affiliation{Department of Physics, Doaba College, Jalandhar 144004, India}

\author{Jiangshan Lan}
\email{jiangshanlan@impcas.ac.cn}
\affiliation{Institute of Modern Physics, Chinese Academy of Sciences, Lanzhou 730000, China}
\affiliation{School of Nuclear Science and Technology, University of Chinese Academy of Sciences, Beijing 100049, China}
\affiliation{Lanzhou University, Lanzhou 730000, China}

\author{Chandan Mondal}
\email{mondal@impcas.ac.cn}
\affiliation{Institute of Modern Physics, Chinese Academy of Sciences, Lanzhou 730000, China}
\affiliation{School of Nuclear Science and Technology, University of Chinese Academy of Sciences, Beijing 100049, China}

\author{Harleen Dahiya}
\email{dahiyah@nitj.ac.in}
\affiliation{Department of Physics, Dr. B. R. Ambedkar National Institute of Technology, Jalandhar 144011, India}

\date{\today}

\begin{abstract}
 We investigate the tomographical structure of pion and kaon in light cone quark model (LCQM). In particular, we study the parton distribution amplitude (PDA) of pion and kaon. We obtain the parton distribution function (PDF) and the generalized parton distributions (GPDs) of the pion and kaon. The valence quark PDA and PDF of pion, after QCD evolution, are found to be consistent with the data from the E791 and the E615 experiments at Fermilab, respectively. Further, we investigate the transverse momentum distributions (TMDs) of pion and kaon. We also discuss the unpolarized TMD evolution for pion and kaon in this model.

\end{abstract}

\maketitle
\section{Introduction}
The nonperturbative structure of hadron is well described by the distribution of partons inside the hadron in both position and momentum space.  The distribution amplitudes (DAs) are among the most basic quantities which not only provide important information on bound states in QCD but also play an essential role in describing the various hard exclusive processes \cite{Brodsky:1989pv,Chernyak:1981zz} of QCD via the factorization theorem \cite{Collins:1996fb} analogous to parton distributions in inclusive processes. DAs are the longitudinal projection of the hadronic wave functions obtained by integrating out the transverse momenta of the constituents of the hadron \cite{Lepage:1980fj,Efremov:1979qk}. The lowest moments of the hadronic DAs for a quark
and an antiquark inside a meson also give us the knowledge of decay constants and transition form factors \cite{ Polyakov-da,tff, tff1, tff2}.
The parton distribution functions (PDFs) \cite{soper, martin}, which are accessible in hard inclusive processes such as deep inelastic scattering (DIS) or Drell-Yan processes,  encode the distribution of longitudinal momentum and polarization carried by
the constituents. The  generalized parton distributions (GPDs) \cite{diehl, garcon, geoke, belitsky} reveal the parton distribution in the direction transverse to the hadron motion providing the spatial distribution. Unlike the PDFs which are function of  longitudinal momentum fraction carried by the active parton $x$ only, GPDs being function of $x$, the longitudinal momentum transferred $\zeta$, and the total momentum transferred from initial state to final state of the hadron $t$, provide  us the knowledge of 3-dimensional (3D) spatial structure of hadron. One can obtain the form factors, charge distributions, PDFs etc. from GPDs under certain conditions \cite{guidal-ff, nikkhoo-ff, miller-charge}.
The momentum tomography of hadrons is described by the transverse momentum-dependent parton distributions (TMDs) \cite{tmd1, tmd2, tmd3}. The TMDs are function of longitudinal momentum fraction $x$ and transverse momentum possessed by the parton ($\textbf{k}_\perp$).

The experimental data for PDFs has been extracted by CTEQ \cite{pdf-1, pdf-11}, NNPDF \cite{pdf-2, pdf-22}, ABM \cite{pdf-3, pdf-33}, GRV/GJR \cite{pdf-4, pdf-44}, MRST/MSTW \cite{martin, pdf-55}, HERAPDF by H1 and ZEUS collaborations \cite{pdf-6}.
The Drell-Yan dilepton production in $\pi^-$-tungsten reactions~\cite{dy-exp3,Sutton:1991ay} is one of the available experiments with access to the pion PDFs. Several next-to-leading order (NLO) analyses of this Drell-Yan process have been studied by Refs.~\cite{Sutton:1991ay,Gluck:1999xe,Wijesooriya:2005ir}. Meanwhile, the reanalysis of the data for the Drell-Yan process including the next-to-leading logarithmic threshold resummation effects have been performed in Ref.~\cite{Aicher:2010cb}.

Experimentally, the internal structure of hadron via GPDs can be extracted from the hard exclusive processes, for example, deeply virtual Compton scattering (DVCS) \cite{dvcs1, dvcs2, dvcs3, dvcs4} and deeply virtual meson production (DVMP) \cite{dvmp1, dvmp2}. GPDs can also be accessed through the time-like Compton scattering \cite{time-like1, time-like2, time-like3}, $\rho$-meson photoproduction \cite{meson-photo-prod, meson-photo-prod1, meson-photo-prod2}, exclusive pion or photon-induced lepton pair-production \cite{lp, lp1, gpd-exp}. The heavy charmonia photoproduction is used to extract the gluon GPDs \cite{charmonia}. The data to access the GPDs of hadrons has been taken from the experiments at J-PARC, Hall-A and Hall-B of JLab with CLAS collaboration and COMPASS at CERN \cite{gpd-exp, gpd-exp1, gpd-exp2, gpd-exp3, gpd-exp4, gpd-exp5, gpd-exp6, gpd-exp7}. The TMDs can be measured through the processes, namely semi-inclusive deep inelastic scattering (SIDIS) \cite{sidis-th, sidis-th1, sidis-th2} and Drell-Yan (DY) process \cite{drell-yan, drell-yan1, drell-yan2, drell-yan3}. The upgraded experiments at JLab, DESY, EIC (electron-ion collider) are valuable in accessing the SIDIS data \cite{sidis, sidis1, sidis2,sidis3} and the data of DY-process are extracted via experiments at J-PARC, BNL, CERN, FNAL  \cite{dy-exp, dy-exp1, dy-exp2, dy-exp3, dy-exp4}.

From the theoretical point of view, the pion DAs have been studied using Chernyak-Zhitnitsky (CZ) \cite{cz} and  Goloskokov-Kroll \cite{gk, gpd-exp} approaches. Further, the pion and the kaon DAs have been theoretically calculated using the Dyson-Schwinger equations \cite{dse, dyson} as well as from Poincar$\acute{\text{e}}$-covariant Bethe-Salpeter wave functions \cite{wf-da-bethe}.
 The pion PDFs have been studied in Nambu-Jona-Lasinio (NJL) model \cite{pdf-njl}. The valence-quark pion and kaon PDFs incorporated with gluon contributions have been studied in Ref. \cite{chen-pdf}. The pion and the kaon DAs and PDFs have been discussed in light-front constituent quark model by using the symmetric quark-bound state vertex function \cite{da-constituent} and chiral constituent quark model \cite{pdf-chiral}. 
The pion PDF has also been the subject of detailed analyses in the phenomenological models in Refs.~\cite{Frederico:1994dx,Shigetani:1993dx,Weigel:1999pc}, also including anti-de Sitter (AdS)/QCD models~\cite{zero-skewness,Gutsche:2013zia,Ahmady:2018,deTeramond:2018ecg} and the chiral quark model~\cite{gpd-chiral}.  The pion PDFs have also been studied within lattice QCD~\cite{Brommel:2006zz,Martinelli:1987bh,Detmold:2003tm,Abdel-Rehim:2015owa,Oehm:2018jvm,Lin:2017snn,Joo:2019bzr,Sufian:2020vzb}.
 Additionally, the first global fit analysis of PDFs in the pion has been reported in Ref.~\cite{Barry:2018ort}.
Although the PDFs are expected to be universal, tension exists regarding the behavior of the pion valence PDF. The large-$x$ behavior of the pion valence PDF is expected to fall off linearly or slightly faster from the analyses of the Drell-Yan data \cite{Sutton:1991ay,Wijesooriya:2005ir}. This is supported by the constituent quark models~\cite{Frederico:1994dx,Shigetani:1993dx}, the Nambu--Jona-Lasinio (NJL) model~\cite{Shigetani:1993dx}, and duality arguments~\cite{Melnitchouk:2002gh}. However, this observation disagrees with perturbative QCD, where the behavior of the same function has been predicted to be ${(1-x)^2}$~\cite{Farrar:1979aw,Berger:1979du,Brodsky:2006hj,Yuan:2003fs}, a behavior further supported by the Bethe-Salpeter equation (BSE) approach~\cite{Hecht:2000xa,Ding:2019lwe}.
Meanwhile, the reanalysis of the data for the Drell-Yan process~\cite{Aicher:2010cb} including the next-to-leading logarithmic threshold resummation effects shows a considerably softer valence PDF at high $x$  when compared to the NLO analysis \cite{Sutton:1991ay,Wijesooriya:2005ir}.
  The pion GPDs have been attempted using covariant and light-front constituent quark models \cite{covariant-n-constituent}, two chiral quark models-NJL model and spectral quark model \cite{gpd-chiral}. The skewed and double quark distributions in the pion using the effective
chiral theory based on the instanton vacuum has been studied in Ref.~\cite{Polyakov:1999gs}. Recently, the GPDs of pion for zero skewedness \cite{zero-skewness} and for non-zero skewedness \cite{non-zero-skewness} have been studied with AdS/QCD approach.  Furthermore, the studies on pion and kaon GPDs are explained in Poincar\'e covariant Bethe-Salpeter constituent quark model by considering the support parameter of pseudoscalar mesons \cite{gpd-pseudo}.

TMDs contain important information on the
3D internal structure of hadrons, especially
the spin-orbit correlations of quarks within them \citep{Angeles-Martinez:2015sea}.  The pion TMDs have been studied by considering the Drell-Yan process with pion beams \cite{tmd-def}. The TMD fragmentation functions of elementary particles in pion and kaon from the NJL-jet model have been explained in Ref. \cite{frag-njl} using Monte-Carlo approach. The pion TMDs are also evaluated in NJL model with Pauli-Villars regularization \cite{tmd-njl}. The transverse structure of the pion in momentum space inspired by light-front holography has been reported in Refs. \cite{tmd-evolution,Ahmady:2019yvo}. A comparative study of the  pion TMDs beyond leading twist in a light-front constituent quark model, the bag model
and a spectator model has been reported in Ref.~\cite{Lorce:2016ugb}.

To understand the relativistic effects of the motion of quarks and gluons in the hadrons,  light-cone formalism is used which is a convenient frame to study the applications to the exclusive processes. The Wigner-rotation is taken into account, when one transforms a composite system from  one reference frame to another. The advantage of using the light-front dynamics is that the Wigner rotation related to the spin states is unity in different frames under the Lorentz transformation. The light-cone quark model (LCQM) finds application in QCD low-scale regime. The pion has chiral symmetry constraints, particularly the explicit chiral and spontaneous symmetry breaking, leading to the pion structure being the simplest valence-quark substructure to study. The LCQM is successful in explaining the electromagnetic form factors of pion and kaon. The results of electromagnetic form factors (EFFs) have already been compared with the experimental data available at low-energy scale in Ref. \cite{pion-1, kaon}, and is found to be consistent with the results evaluated in this model. The decay constants and charge radii for both particles have also been predicted in the LCQM. In light of the progress, it therefore becomes necessary to enhance this model to study several distributions of partons in the mesons.

In the present work,  we have implemented the Melosh-Wigner transformation to derive the light-cone spin-flavor wave functions of pion and kaon. We have investigated the DAs of quark in pion and kaon at the model initial energy scale $\mu_0$. The QCD evolution is applied on the low-energy scale model calculated DAs and compared with asymptotic result. The moments at different evolution scales corresponding to pion and kaon are also compared with different available theoretical predictions. Further, we evaluate the PDFs for both pseudoscalar mesons in this model. To compare the model results with the available experimental data of PDFs, the next-to-next-leading order (NNLO) DGLAP evolution is put into consideration by choosing the appropriate scales. We have studied the GPDs and the TMDs of pion and kaon from the overlap of light-cone wave functions (LCWFs). Spin-0 hadrons refer to two GPDs. The chirally even GPD, $H(x, \zeta, t)$ describes the distribution for an unpolarized quark, whereas the chirally-odd GPD, $E_T(x, \zeta, t)$ corresponds to the distribution of a transversely polarized quark inside the hadron \cite{spin-structure}. Meanwhile, there are two leading-twist TMDs in case of the pseudoscalar meson: the unpolarized quark TMD, $ f_1(x,{\bf k}^2_\perp)$ and the transversely-polarized quark TMD, $h_1^\perp(x,{\bf k}^2_\perp)$, also known as Boer-Mulders function~\cite{tmd2,tmd3}. $ f_1(x,{\bf k}^2_\perp)$ is a T-even distribution, whereas $h_1^\perp(x,{\bf k}^2_\perp)$ is naively a T-odd distribution and such distribution is dynamically generated by initial or final state interactions~\cite{Brodsky:2002cx,Brodsky:2002rv}. Here, we shall use the perturbative gluon rescattering in order to generate the Boer-Mulders function for light mesons.

The manuscript is organized as follows. In Section II, we discuss the details of light-cone quark model. The wave functions and distribution amplitudes for pion and kaon are detailed in Section III. Section IV includes the parton distribution functions for both pion and kaon mesons and its DGLAP evolution to the higher energy scales. In Section V, a detailed description and graphical interpretations of the GPDs are given for $u$-quark in pion and kaon. In Section VI, we present the quark TMDs of both pseudoscalar mesons. In this section, the details of the scale evolution of the unpolarized quark TMD is also presented. Finally, the results are concluded in Section VII.

\section{II Light-cone quark model (LCQM)}
The hadron eigenstate $|M (P^+, \textbf{P}_\perp, S_z)\rangle $ in connection with multi-particle Fock eigenstates $|n \rangle$ is defined as \cite{meson-state}
\begin{widetext}
\begin{eqnarray}
|M (P^+, \mathbf{P}_\perp, S_z) \rangle
   =\sum_{n,\lambda_i}\int\prod_{i=1}^n \frac{\mathrm{d} x_i \mathrm{d}^2
        \mathbf{k}_{\perp i}}{\sqrt{x_i}~16\pi^3}
 16\pi^3  \delta\Big(1-\sum_{i=1}^n x_i\Big)
\delta^{(2)}\Big(\sum_{i=1}^n \mathbf{k}_{\perp i}\Big) \psi_{n/M}(x_i,\mathbf{k}_{\perp i},\lambda_i)   | n ; x_i P^+, x_i \mathbf{P}_\perp+\mathbf{k}_{\perp i},
        \lambda_i \rangle
        ,\nonumber\\
\end{eqnarray}
\end{widetext}
where $P=(P^+,P^-,\textbf{P}_\perp)$ is considered as the total momentum of meson and $S_z$ is the longitudinal spin projection. The momenta of meson having mass $M$ and its constituents having masses $m_1$ and $m_2$ in light-cone frame are defined as
\begin{eqnarray}
P&=&\bigg(P^+,\frac{M^2}{P^+},\textbf{0}_\perp \bigg),\\
k_1&=&\bigg(x P^+, \frac{\textbf{k}_\perp^2+m_1^2}{x P^+},\textbf{k}_\perp \bigg),\\
k_2&=&\bigg((1-x) P^+, \frac{\textbf{k}_\perp^2+m_2^2}{(1-x) P^+},-\textbf{k}_\perp \bigg).
\end{eqnarray}
The multi-particle Fock states containing $n$ constituents where the $ith$ constituent holding the light-cone longitudinal momentum fraction $x_i=\frac{k_i^+}{P^+}$, the transverse momentum $\textbf{k}_{\perp i}$ and  helicity $\lambda_i$ are normalized as
\begin{eqnarray}
&&\langle{n;{k'}_i^+, {\textbf{k}'}_{\perp i},\lambda'_i} | {n;{k}_i^+, {\textbf{k}}_{\perp i},\lambda_i}\rangle\nonumber\\
&=&\prod_{i=1}^n 16 \pi^3 k_i^+ \delta({k'}_i^+ -k_i^+) \delta^{(2)} ({\textbf{k}'}_{\perp i}-{\textbf{k}'}_{\perp i})\delta_{\lambda'_i \lambda_i}.
\end{eqnarray}
The light-cone wave function in LCQM is written as
\begin{eqnarray}
\psi_{S_z}^F(x,\textbf{k}_\perp, \lambda_1, \lambda_2)=\varphi(x, \textbf{k}_\perp)\ \chi_{S_z}^F(x,\textbf{k}_\perp, \lambda_1, \lambda_2).
\label{space}
\end{eqnarray}
where $\varphi$ and $\chi$ correspond to the momentum space and spin wave functions respectively and superscript $F$ stands for the front form.

The LCWF of pion (or kaon) can be obtained through the transformation of the instant-form SU(6) wave functions using Melosh-Wigner rotation. The spin wave function of the pseudoscalar meson in the instant form ($T$) can be written as \cite{pion, kaon}
\begin{eqnarray}
\chi_T=\frac{(\chi^{\uparrow}_1 \chi^{\downarrow}_2-\chi^{\uparrow}_2 \chi^{\downarrow}_1)}{\sqrt{2}},
\label{spin}
\end{eqnarray}
where $\chi^{\uparrow,\downarrow}_i$ is the two-component Pauli spinor. One can relate the light-cone spin states $|J,\lambda \rangle_F$ and the ordinary instant-form spin states $|J,s \rangle_T$ as follows
\begin{eqnarray}
|J,\lambda \rangle_F=\sum_s U^J_{s\lambda} |J,s \rangle_T,
\label{melosh}
\end{eqnarray}
where $U^J$ is the Melosh-Wigner rotation operator.

The spin space wave function of pseudoscalar meson can obtained in the infinite momentum frame by implementing the transformation Eq. (\ref{melosh}) in Eq. (\ref{spin}). The Melosh-Wigner transformation is used to connect the instant form spin states and light-front form spin states as
\begin{eqnarray}
\chi_i^\uparrow(T)&=&\omega_i[(q_i^+ +m_i)\chi_i^\uparrow(F)-q_i^R \chi_i^\downarrow(F)],\label{instant-front1}\\
\chi_i^\downarrow(T)&=&\omega_i[(q_i^+ +m_i)\chi_i^\downarrow(F)+q_i^L\chi_i^\uparrow(F)].
\label{instant-front}
\end{eqnarray}
Here we take the instant-form four-momenta for 2 quarks as: $q_1^\mu =(q_1^0,\textbf{q})$ and $q_2^\mu=(q_2^0,-\textbf{q})$ with $q_i^0=(m_i^2+\textbf{q}^2)^{1/2}.$  In Eqs. (\ref{instant-front1}) and (\ref{instant-front}), $\omega_i=[2q_i^+(q_i^0+m_i)]^{1/2}$ and $q_i^{R,L}=q_i^1 \pm i q_i^2$. A meson is a bound state of a quark ($Q$) and an antiquark ($\bar{Q}$) viz. $Q\bar{Q}$, where the masses of two partons are denoted as $m_1$ and $m_2$. For pion, we consider $m_1=m_2=m$ whereas for kaon, we take $m_1 \neq m_2$ because of its composition.

The light-cone spin wave function of pseudoscalar `$\mathcal{P}$' meson, which can be pion or kaon depending on their composition, has the form
\begin{eqnarray}
\chi^\mathcal{P}(x,\textbf{k}_\perp)=\sum_{\lambda_1, \lambda_2}\kappa_{S_z}^F(x,\textbf{k}_\perp,\lambda_1, \lambda_2) \chi_1^{\lambda_1}(F) \chi_2^{\lambda_2}(F),
\end{eqnarray}
with $S_z$ and $\lambda$ being the spin projection of pion (or kaon) and quark helicity, respectively.
Since for pion (kaon) having masses $m$ ($m_1 \ \rm{and}\  m_2$), the $z$-component of spin is zero ($S_z=0$), therefore, the component coefficients $\kappa_{S_z=0}^F(x,\textbf{k}_\perp,\lambda_1, \lambda_2)$ in spin wave function are indicated as
\begin{eqnarray}
\kappa_0^F (x, \textbf{k}_\perp, \uparrow, \downarrow)&=&  [(q_1^+ +m_1)\nonumber\\&&\times(q_2^+ + m_2)-q_\perp^2]/\sqrt{2}\, \omega_1 \omega_2, \nonumber\\
\kappa_0^F (x, \textbf{k}_\perp, \downarrow, \uparrow)&=&  - [(q_1^+ +m_1)\nonumber\\&&\times(q_2^+ + m_2)-q_\perp^2]/\sqrt{2}\, \omega_1 \omega_2, \nonumber\\
\kappa_0^F (x, \textbf{k}_\perp, \uparrow, \uparrow)&=&   [(q_1^+ +m_1)q_2^L\nonumber\\&&-(q_2^+ + m_2)q_1^L]/\sqrt{2}\, \omega_1 \omega_2, \nonumber\\
\kappa_0^F (x, \textbf{k}_\perp, \downarrow, \downarrow)&=& [(q_1^+ +m_1)q_2^R\nonumber\\&&-(q_2^+ + m_2)q_1^R]/\sqrt{2}\, \omega_1 \omega_2, \label{coeff}
\end{eqnarray}
where $q_1^+=q_1^0+q_1^3=x_1 \mathcal{M}$, $q_2^+=q_2^0+q_2^3=x_2 \mathcal{M}$, and $\textbf{k}_\perp=\textbf{q}_\perp$, with
\begin{eqnarray}
\mathcal{M}^2=\frac{m_1^2+\textbf{k}^2_\perp}{x_1}+\frac{m_2^2+\textbf{k}^2_\perp}{x_2}.
\label{m}
 \end{eqnarray}
Here $x_i$ ($i=1,2$) is the light-cone quark momentum fraction in light-front dynamics with the constraint
\begin{eqnarray}
\sum_{i=1}^n x_i=1.
\end{eqnarray}
This leads to $x_1+x_2=1$ for the case of meson. If we assume the momentum fraction of one parton $x_1=x$, then for the other parton it becomes $x_2=1-x$.

The component coefficients $\kappa_{S_z=0}^F(x,\textbf{k}_\perp,\lambda_1, \lambda_2)$ in spin wave function given in Eq. (\ref{coeff}) must satisfy the following normalization conditions for pion (or kaon)
\begin{eqnarray}
\sum_{\lambda_1,\lambda_2} \kappa_0^{F*}(x, \textbf{k}_\perp, \lambda_1, \lambda_2)\kappa_0^F(x, \textbf{k}_\perp, \lambda_1, \lambda_2)=1.
\end{eqnarray}

The momentum space wave functions $\varphi^{\pi(K)}(x,\textbf{k}_\perp)$ in Eq. (\ref{space}) are adopted using Brodsky-Huang-Lepage (BHL) prescription. For pion we have,
\begin{eqnarray}
\varphi^\pi(x,\textbf{k}_\perp)=A^\pi \ {\rm exp}\bigg[-\frac{1}{8 \beta_\pi^2} \frac{{\bf k}^2_\perp+m^2} {x(1-x)} \bigg],
\label{bhl-pi}
\end{eqnarray}
and for kaon we have
\begin{eqnarray}
\varphi^K(x,\textbf{k}_\perp)&=& A^K \ {\rm exp} \Bigg[-\frac{\frac{\textbf{k}^2_\perp+m_1^2}{x}+\frac{\textbf{k}^2_\perp+m_2^2}{1-x}}{8 \beta_K^2} \nonumber\\
&-&\frac{(m_1^2-m_2^2)^2}{8 \beta_K^2 \bigg(\frac{\textbf{k}^2_\perp+m_1^2}{x}+\frac{\textbf{k}^2_\perp+m_2^2}{1-x}\bigg)}\Bigg],
\label{bhl-k}
\end{eqnarray}
where $A^\pi$ and $A^K$ are the normalization constants for pion and kaon, respectively.

The two-particle Fock state expansion can be described in terms of LCWFs $\psi_{S_z}(x, \textbf{k}_\perp, \lambda_1, \lambda_2)$ defined in Eq. (\ref{space}) and we have
\begin{widetext}
\begin{eqnarray}
\ket{\pi(K)(P^+,\textbf{P}_\perp,S_z)}&=&\int \frac{{\rm d}^2\textbf{k}_\perp  {\rm d}x}{\sqrt{x(1-x)}2 (2 \pi)^3}\big[\psi^{\pi(K)}_{S_z}(x,\textbf{k}_\perp, \uparrow, \uparrow)\ket{x P^+, \textbf{k}_\perp, \uparrow, \uparrow} +\psi^{\pi(K)}_{S_z}(x,\textbf{k}_\perp, \uparrow, \downarrow)\ket{x P^+, \textbf{k}_\perp, \uparrow, \downarrow}\nonumber\\
&&+\psi^{\pi(K)}_{S_z}(x,\textbf{k}_\perp, \downarrow, \uparrow)\ket{x P^+, \textbf{k}_\perp, \downarrow, \uparrow}+\psi^{\pi(K)}_{S_z}(x,\textbf{k}_\perp, \downarrow, \downarrow)\ket{x P^+, \textbf{k}_\perp, \downarrow, \downarrow}\big].
\label{overlap}
\end{eqnarray}
\end{widetext}
\section{III wave functions and Distribution amplitudes (DAs)}
\begin{figure}
\centering
\includegraphics[width=.42\textwidth]{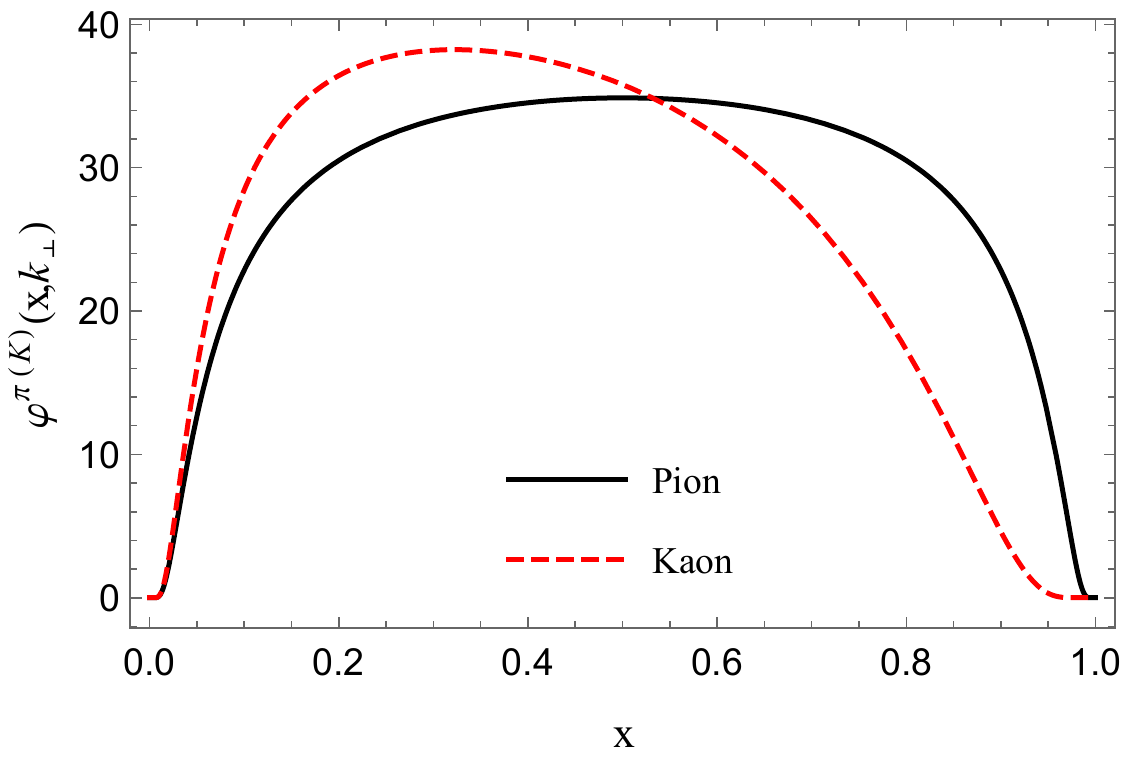}
\caption{The solid black curve represents the wave function $\varphi^\pi (x,k_\perp)$ for pion and dashed red curve represents the kaon wave function $\varphi^K(x,k_\perp)$ as a function of $x$ with fixed $\textbf{k}_\perp = 0.2~ \rm GeV$.}
\label{wf}
\end{figure}

The constituent quark masses and the harmonic scale $\beta$ are only two input parameters required to compute the pion and the kaon distribution funtions. The parameters used in the present work are listed in Table \ref{parameters}. Comparison between the momentum space wavefuntion of the pion $\varphi^\pi(x,\textbf{k}_\perp)$ and the kaon $\varphi^K(x,\textbf{k}_\perp)$ is shown in Fig. \ref{wf}. The pion wave function shows symmetry over $x=0.5$ whereas, due to dissimilar quark masses, kaon wave function appears to be asymmetrical along $x$.

\begin{table}[ht]
\begin{center}
\begin{tabular}{|@{\hspace{18pt}} c @{\hspace{12pt}} |
@{\hspace{12pt}} c @{\hspace{12pt}} | @{\hspace{12pt}} c
@{\hspace{12pt}} |  }

\hline
Meson & {Mass in GeV {\hspace{9pt}} } & {$\beta$ in $\rm GeV$ {\hspace{9pt}}} \\ \hline\hline
$\pi$ $(u\overline{d})$ & $m= 0.2$ & $0.410$ \\
$K$ $(u\overline{s})$ & $m_1=0.2$, $m_2=0.556$ & $0.405$ \\
\hline
\end{tabular}
\caption{The valence quark masses and harmonic scale $\beta$ parameters in the pion and the kaon.}
\label{parameters}
\end{center}
\end{table}

LCWFs give unique access to light cone distributions
by integrating out the transverse momentum \cite{Lepage:1980fj}. Among those, the DAs control the exclusive processes at large momentum transfer. In
light-front formalism, the leading-twist DAs for pseudoscalar mesons are defined through the correlation \cite{Bodwin:2006dm,Braguta:2006wr,Braguta:2007fh,da_evolution}
\begin{eqnarray}
&&\bra{0}\bar{\Psi}(z)\gamma^+ \gamma_5\Psi(-z)\ket{\mathcal{P}^+(P)}\nonumber\\
&&=ik^+ f_{\mathcal{P}} \int_0^1 dx e^{i (x-1/2) k^+ z^-}\phi(x)\Big\vert_{z^+,{\bf z}_\perp=0},
\end{eqnarray}
where $\mathcal{P}$ represents pseudoscalar meson and the decay constant is denoted with $f_\mathcal{P}$.
Substituting the pseudoscalar meson states and the quark field operators lead to the definition of DAs in terms of LCWFs, one has \cite{da_evolution,Choi:2007yu}
\begin{eqnarray}
\frac{f_{\pi(K)}}{2\sqrt{2 N_c}}\phi(x)&=&\frac{1}{\sqrt{2 x(1-x)}} \int \frac{d^2 {\bf k}_\perp}{16 \pi^3}\Big[\psi^{\pi(K)}_0(x, {\bf k}_\perp, \uparrow ,\downarrow)\nonumber\\&&-\psi^{\pi(K)}_0(x, {\bf k}_\perp,\downarrow,\uparrow)\Big],
\end{eqnarray}
with the normalization condition at any scale:
\begin{eqnarray}
\int_0^1 dx  \phi(x, \mu)=1.
\end{eqnarray}
Using the LCWFs given in Eq.~(\ref{space}), we compute the DAs of the pion and the kaon at the model scale. Next, the LO QCD evolution of the DAs is carried using the Efremov-Radyushkin-Brodsky-Lepage (ERBL) equations \cite{Efremov:1979qk,Lepage:1980fj}. In a Gegenbauer basis, one has \cite{da_evolution_2}

\begin{eqnarray}
\phi^{\pi(K)}(x,\mu)=6x (1-x)\sum_{n=0}^\infty C_n^{\frac{3}{2}}(2x-1) a_n(\mu),
\end{eqnarray}
with
\begin{eqnarray}
a_n(\mu)&=&\frac{2(2n+3)}{3(n+1)(n+2)}\bigg(\frac{\alpha_s(\mu)}{\alpha_s(\mu_0)}\bigg)^{\frac{\gamma^{(0)}_n}{2\beta_0}}\nonumber\\ &\times&\int_0^1 dx C_n^{\frac{3}{2}}(2x-1)\phi^{\pi(K)}(x,\mu_0),
\end{eqnarray}
where $C_n^{\frac{3}{2}}(2x-1)$ is a Gegenbauer polynomial.
The strong coupling constant $\alpha_s(\mu)$ is given by
\begin{eqnarray}
\alpha_s(\mu)=\frac{4\pi}{\beta_0 ~\ln(\frac{\mu^2}{\Lambda^2_{QCD}})}.
\end{eqnarray}
The factor $\frac{\gamma^{(0)}_n}{2\beta_0}$ defines the anomalous dimensions
\begin{eqnarray}
\gamma^{(0)}_n=-2 c_F\bigg(3+\frac{2}{(n+1)(n+2)}-4 \sum_{m=1}^{n+1}\frac{1}{m}\bigg),
\end{eqnarray}
and
\begin{eqnarray}
\beta_0=\frac{11}{3}c_A-\frac{2}{3}n_F,
\end{eqnarray}
where $c_A=3$ and $n_F$ correspond to the number of active flavors. The color factor $c_F=\frac{4}{3}$ and  $\Lambda_{\mathrm{QCD}}=0.226$ GeV \cite{Gutsche:2013zia}.

\begin{figure*}
\centering
\begin{minipage}[c]{1\textwidth}
(a)\includegraphics[width=.42\textwidth]{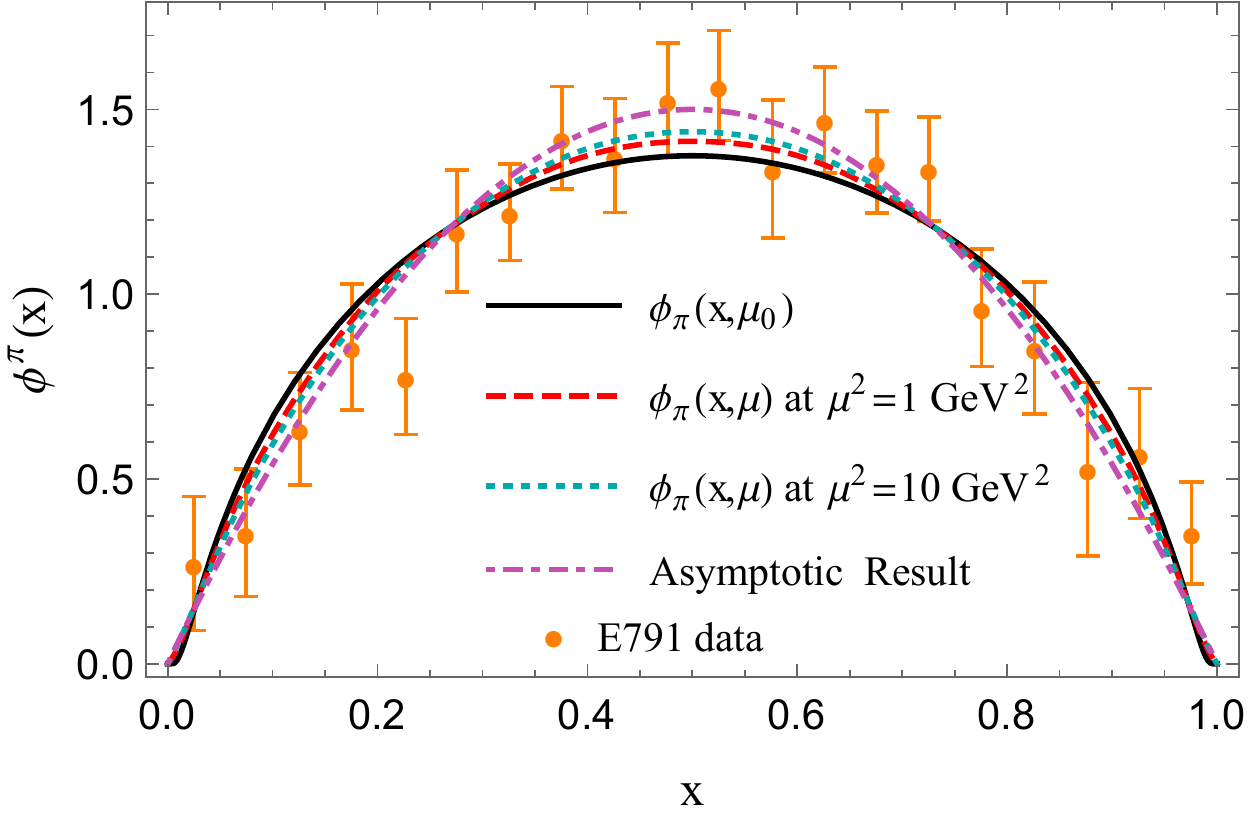}
(b)\includegraphics[width=.42\textwidth]{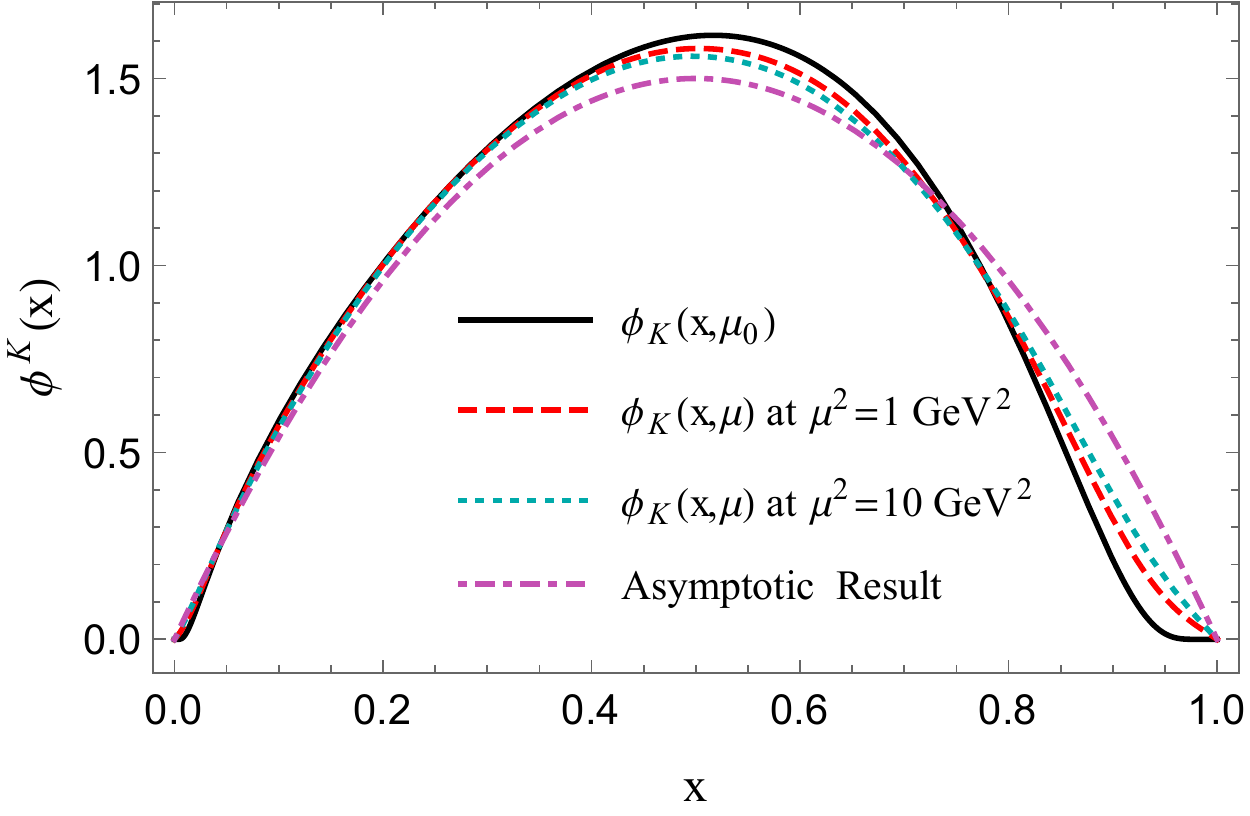}
\end{minipage}
\caption{(a) Left panel : pion DA at initial scale $\mu^2_0=0.246$ $\rm GeV^2$ (solid black curve), which is evolved to $\mu^2=1 $ $\rm GeV^2$ (dashed red curve) and $\mu^2=10 $ $\rm GeV^2$ (dotted cyan curve). The model results are compared with the asymptotic result $(\phi(x)=6x(1-x))$ (dot-dashed purple curve) and the experimental data of E791 (orange data points) \cite{Aitala:2000hb}. (b) Right panel : the kaon DA at initial scale (solid black curve), which is evolved to $\mu^2=1 $ $\rm GeV^2$ (dashed red curve) and $\mu^2=10 $ $\rm GeV^2$ (dotted cyan curve) and compared with the asymptotic result (dot-dashed purple curve).}
\label{da}
\end{figure*}

In Fig.~\ref{da}(a) and fig.~\ref{da}(b), we show the evolution of the pion and the kaon DAs, respectively from the initial scales $\mu_0^2$ to $\mu^2=10$ GeV$^2$ which is the scale relevant to the E791 data \cite{Aitala:2000hb}. As can be seen, the pion DA is close to the asymptotic DA already at $\mu^2=1$ GeV$^2$ and the DA approaches towards the asymptotic DA with increasing evolution scale, however, the effect is small. The evolved pion DA in this model shows a good agreement with the E791 data. Unlike the pion,
the evolved kaon DA is still distinct from the
asymptotic DA at $\mu^2=1$ GeV$^2$ or even at $\mu^2=10$ GeV$^2$. However, the trend of evolution is same as the pion DA.
\begin{table*}
  \centering
  \begin{tabular}{|c|c|c|c|c|}
    \hline
 Pion DA   & $\mu$ [GeV] &$\langle \xi_2 \rangle$ &$\langle \xi_4 \rangle$&$\langle x^{-1} \rangle$ \\
     \hline\hline

     Asymptotic & $\infty$  & $0.200$ &$0.085$ & $3.00$ \\


     LCQM (This work) & $1,2$  & $0.212, 0.21$ &$0.094, 0.092$ & $3.05, 3.05$ \\

    LF Holographic ($B=0$) \cite{Ahmady:2018} &$1,2$  & $0.180,0.185$ &$0.067,0.071$ &$2.81,2.85 $\\

    LF Holographic ($B \gg 1$) \cite{Ahmady:2018} &$1,2$  & $0.200,0.200$ &$0.085,0.085$&$2.93,2.95 $\\

    LF Holographic \cite{Brodsky:2007hb} &$ \sim 1$  & $0.237$ &$0.114$&$4.0$\\

         Platykurtic \cite{Stefanis:2014nla} & $2$ & $0.220^{+0.009}_{-0.006}$& $0.098^{+0.008}_{-0.005}$ & $3.13^{+0.14}_{-0.10}$\\

    LF Quark Model \cite{Choi:2007yu} & $\sim 1$ & $0.24 [0.22]$ &$0.11 [0.09]$ & \\

    Sum Rules \cite{Ball:2004ye} & $1$ & $0.24$ & $0.11$&\\

       Renormalon model \cite{Agaev:2005rc}& $1$ & $0.28$ & $0.13$&\\

     Instanton  vacuum \cite{Petrov:1998kg,Nam:2006au}  & $1$ & $0.22,0.21$ & $0.10,0.09$&\\

          NLC Sum Rules \cite{Bakulev:2001pa} &$2$ &$0.248^{+0.016}_{-0.015}$  & $0.108^{+0.05}_{-0.03}$ &$3.16^{+0.09}_{-0.09}$ \\

    Sum Rules\cite{cz}& $2$ & $0.343$ &$0.181$ &$4.25$\\

    Dyson-Schwinger[RL,DB]\cite{Chang:2013pq} & $2$ & $0.280,0.251$ &$0.151,0.128$ &$5.5,4.6$\\

    Lattice \cite{Arthur:2010xf} &$2$& $0.28(1)(2)$& & \\

      Lattice \cite{Braun:2015axa} & $2$ & $0.2361(41)(39)$ & &\\

     Lattice \cite{Braun:2006dg} & $2$ & $0.27 \pm 0.04$ & &\\
     \hline
    \end{tabular}
  \caption{Comparison of first two possible moments and inverse moment in this model with the available theoretical results for pionic DA.}
  \label{tab:PionDAmoments}
\end{table*}

\begin{table*}
  \centering
  \begin{tabular}{|c|c|c|c|c|c|c|}
    \hline
  Kaon DA   & $\mu$ [GeV] &$\langle \xi_1 \rangle$ & $\langle \xi_2 \rangle$&$\langle \xi_3 \rangle$ &$\langle \xi_4 \rangle$&$\langle x^{-1} \rangle$ \\
     \hline \hline

     Asymptotic & $\infty$ &0&$0.200$ &0&$0.085$ & $3.00$ \\


     LCQM (This work) & $ 1,2$ & $0.033,0.028$  &$0.183, 0.187$ & $0.019,0.016$ & $0.073, 0.076$ & $3.027,3.037$\\

  LF  Holographic ($B=0$) \cite{Ahmady:2018} &$1,2$  &$0.055,0.047$ &$0.175,0.180$ &$0.021,0.018$ & $0.062,0.067$ &$2.55,2.62$\\

   LF Holographic ($B \gg 1$) \cite{Ahmady:2018} &$1,2$&$0.094,0.081$ &  $0.194,0.195$ &$0.039,0.034$&$0.080,0.081$ & $2.60,2.66$\\

   Lattice \cite{Arthur:2010xf} &$2$&$0.036(2)$&$0.26(2)$&&& \\

   LF Quark Model \cite{Choi:2007yu} &$\sim 1$&$0.06[0.08]$&$0.21[0.19]$&$0.03[0.04]$&$0.09[0.08]$& \\

   Sum Rules \cite{Ball:2006wn} &$1$&$0.036$&$0.286$&$0.015$&$0.143$&$3.57$\\

   Dyson-Schwinger[RL,DB] \cite{Shi:2014uwa}& $2$ & $0.11,0.040$ & $0.24,0.23$ & $0.064, 0.021$ & $0.12, 0.11$ & \\

   Instanton vacuum \cite{Nam:2006au}&$1$&$0.057$&$0.182$&$0.023$&$0.070$&\\
   \hline
  \end{tabular}
  \caption{Comparison of first four possible moments and inverse moment in this model with the available theoretical results for kaonic DA.}
  \label{tab:KDAmoments}
\end{table*}

It is useful to compute the moments in order to
quantitatively compare with other theoretical predictions. The $n$-th
moment is defined as,
\begin{eqnarray}
\langle z_n \rangle = \int_0^1 dx~z^n~\phi(x,\mu),
\end{eqnarray}
where $z$ can be $\xi=(2x-1)$ or $x^{-1}$.
Our predictions for the moments of the pion
DA are compared to other theoretical approaches in
Table \ref{tab:PionDAmoments}. From the table, it is important to note that the LCQM DAs have their moments larger than their asymptotic values in case of pion, and the value goes on decreasing when evolve towards the latter from below.
 LCQM moments show a similar behavior as the moments obtained in other theoretical approaches except LF holographic model, and our moments for the pion DA are closer to the asymptotic results compared to other approaches.
The moments of the kaon DA are compared to other theoretical predictions in
Table \ref{tab:KDAmoments}. In this case, the even moments in LCQM model are lower than their asymptotic values, while
the predicted odd moments are greater than zero, which is their asymptotic value. Further, the inverse moment shows the higher value compared to the asymptotic one. We notice a similar trend as observed in LF holograpic model \citep{Ahmady:2018} where the exception lies in inverse moment.
\section{IV Parton distribution functions (PDFs)}
\begin{figure}
\centering
\includegraphics[width=.42\textwidth]{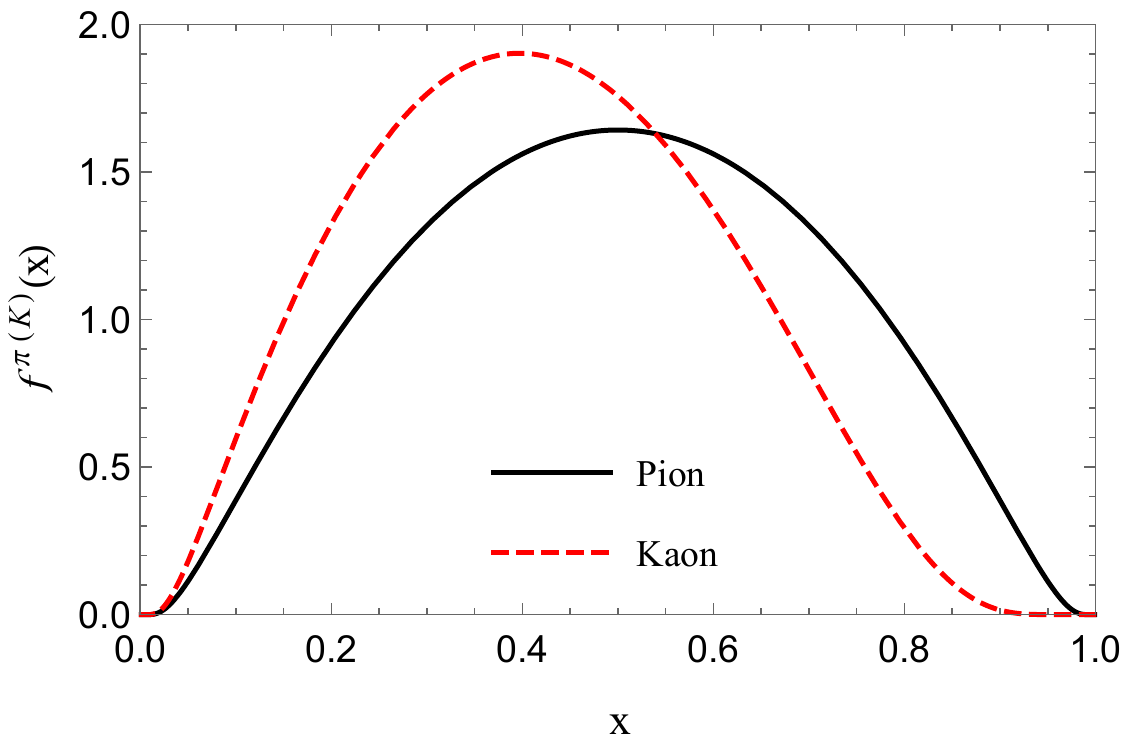}
\caption{ The black solid curve represents the PDF $f^\pi(x)$ in the pion and the red dashed curve represents the kaon PDF $f^K(x)$ for $u$-quark at the model scale.}
\label{pdf}
\end{figure}

The pion (kaon) PDF gives the probablity of finding the quark in the pion (kaon) where the quark carries a longitudinal momentum fraction $x=k^+/P^+$.  At fixed light-front time, the PDF can be expressed as \cite{pdf-def}
\begin{eqnarray}
f^{\mathcal{P}}(x)&=&\frac{1}{2}\int \frac{dz^-}{4 \pi} e^{ik^+ z^-/2}\\&\times &\bra{\mathcal{P}^+(P);S}\bar{\Psi}(0)\Gamma \Psi(z^-)\ket{\mathcal{P}^+(P);S}\vert_{z^+=\textbf{z}_\perp=0}.\nonumber
\end{eqnarray}
Since the spin is zero in both the cases $S=0$, we deal with unpolarized parton distribution function which comes from the above relation by substituting $\Gamma=\gamma^+$.
The overlap form of PDF is defined by putting the pion (kaon) states from Eq. (\ref{overlap}). We have
\begin{eqnarray}
f^{\pi(K)}(x)&=& \int \frac{d^2\textbf{k}_\perp}{16\pi^3} \big[\mid{ \psi_0^{\pi(K)}(x,\textbf{k}_\perp, \uparrow, \uparrow )}\mid^2 \nonumber\\
&&+ \mid \psi_0^{\pi(K)}(x,\textbf{k}_\perp, \uparrow, \downarrow)\mid^2 
\nonumber\\
&& + \mid \psi_0^{\pi(K)}(x,\textbf{k}_\perp, \downarrow, \uparrow )\mid^2
\nonumber\\
&&+\mid \psi_0^{\pi(K)}(x,\textbf{k}_\perp, \downarrow, \downarrow)\mid^2 \big].\nonumber\\
\label{pdf-eq}
\end{eqnarray}
Using the LCWFs given in Eq. (\ref{space}), we evaluate  the quark distribution functions, $f^{\pi(K)}(x)$ at the initial scale and plot them as a function of $x$ in Fig. \ref{pdf}. Due to equal constituent quark (antiquark) mass, the distribution in the pion appears to be symmetric over $x=0.5$, while in kaon, the light quark distribution is maximum at a slightly lower value of quark momentum fraction in the longitudinal direction. The peak is broader in the case of the pion as compared to the kaon. The distribution peak has higher amplitude in case of the kaon as compared to the pion.

We now have our PDFs for the light mesons at scales relevant to constituent quark masses which are several hundred $\mathrm{MeV}$. At the model scales, both PDFs for the valence quark (antiquark) is normalized to $1$:
\begin{equation}
\int_{0}^{1}f(x)\,dx = \int_{0}^{1}f(1-x)\,dx=1.\label{eq:initial_PDF_normalization}
\end{equation}
Meanwhile, within the two-body approximation one can write the momentum sum rule:
\begin{equation}
\int_{0}^{1}x\,f(x)\,dx +\int_{0}^{1}x\,f(1-x)\,dx=1.\label{eq:momentum_sum}
\end{equation}
This states that the valence quark and antiquark together carry the entire light-front momentum of the meson, which is appropriate to a low-resolution model.

\subsection{(A) QCD Evolution for Pion PDF}
\begin{figure}
\centering
\includegraphics[width=.42\textwidth]{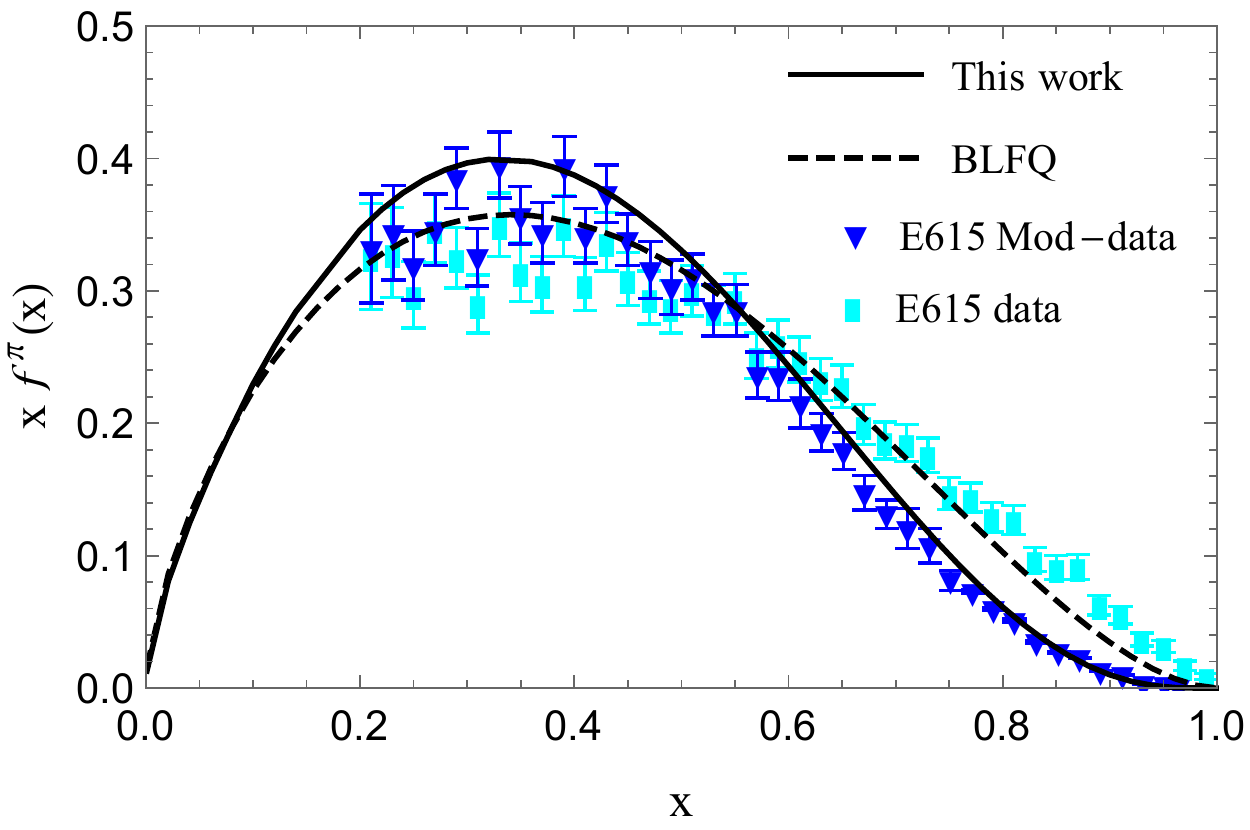}
\caption{The QCD evolution for the pion PDF in LCQM from the initial scale $\mu_0^2=0.246$ GeV$^{2}$ evolved to $\mu^2=16$ GeV$^{2}$ (solid black curve) compared with FNAL-E615 experimental data (cyan data points) \cite{dy-exp4} and modified FNAL-E615 data (blue data points) \cite{Aicher:2010cb}. The dashed black curve denotes the valence quark PDF obtained from BLFQ \cite{blfq}.}
\label{qcd-evolution}
\end{figure}
The valence quark distributions at high $\mu^2$ scale can be determined with the initial input by performing the QCD evolution. We adopt the NNLO DGLAP equations~\cite{Dokshitzer:1977sg,Gribov:1972ri,Altarelli:1977zs} of QCD, to evolve our PDFs from our model scales to higher scales $\mu^2$ needed for the comparison with experiment. The scale evolution allows quarks to emit and absorb gluons while the emitted gluons allowed to generate quark-antiquark pairs as well as additional gluons. In this picture, the higher scale reveals the gluon and sea quark  components of the constituent quarks through QCD.

We explicitly evolve our initial PDFs from the LCQM model for the pion to the relevant experimental scales ${\mu^2=16~\mathrm{GeV}^2}$ using the Higher Order Perturbative Parton Evolution toolkit to numerically solve the NNLO DGLAP equations~\cite{Salam:2008qg}. While applying the DGLAP equations numerically, we impose the condition that the running coupling $\alpha_s(\mu^2)$ saturates in the infrared at a cutoff value of max $\alpha_s$ = 1 \citep{blfq,Lan:2019rba,Lan:2019img,Xu:2019xhk}.

In Fig. \ref{qcd-evolution}, we compare our result for the valence quark PDF of the pion with the FNAL-E-0615 \cite{dy-exp4} and FNAL-E-0615 modified data \cite{Aicher:2010cb}. Also, the theoretical result of the valence quark PDF evaluated by using the basis light front quantization (BLFQ) \cite{blfq} is taken for comparison with our result. We notice that our result does not fit to FNAL-E-0615 experiment. Meanwhile, our result fits to the modified E615 data
after being reanalyzed to take into account soft gluon
resummation \citep{Aicher:2010cb}. We are able to fit the reanalyzed data
using an initial scale $\mu_0^2=0.246$ GeV$^{2}$ (similar to the initial
scale used in Refs~\citep{Gutsche:2013zia,Ahmady:2018,da_evolution_2}). At the initial scale, this model consists only valence quarks, no sea quark or gluon contributes. The behavior of the pion PDF at large $x$ is still an unresolved issue. However,  our observation at large $x$ agrees with perturbative QCD, where the behavior of the PDF has been predicted to be ${(1-x)^2}$~\cite{Farrar:1979aw,Berger:1979du,Brodsky:2006hj,Yuan:2003fs}, a behavior further supported by the Bethe-Salpeter equation (BSE) approach~\cite{Hecht:2000xa,Ding:2019lwe}.
\section{V Generalized parton distributions (GPDs)}
We calculate the GPDs of quark for pion and kaon in LCQM. GPDs have support region $x \epsilon [-1,1]$ \cite{diehl, x-region-brodsky}. However, for present calculations we restrict ourself to only DGLAP region i.e. $\zeta<x<1$.
At leading twist, there are two independent GPDs for spin zero hadrons. One of them is chirally-even and the other is chirally-odd. 
The correlation to evaluate chiral-even GPD, $H(x,\zeta=0,t)$, which corresponds to unpolarized quark in unpolarized meson, is defined through the bilocal operator of light-front correlation functions of the vector current \cite{diehl}
\begin{eqnarray}
H^{\mathcal{P}}(x,0,t)&=& \int \frac{dz^-}{4 \pi} e^{i x P^+ z^-/2}\nonumber\\
&&\bra{\mathcal{P}^+(P')}\bar{\Psi}(0)\gamma^+ \Psi(z)\ket{\mathcal{P}^+(P)}\vert_{z^+=\textbf{z}_\perp=0},\nonumber\\
\end{eqnarray}
while the chiral-odd GPD, $E_{T}(x,\zeta=0,t)$ corresponding to transversely-polarized quark in unpolarized meson is defined through the correlation functions of the tensor current
\begin{eqnarray}
&&\frac{i \epsilon^{ij}_\perp q^i_\perp}{M_\mathcal{P}}E_T^\mathcal{P}(x,0,t)= \int \frac{dz^-}{4 \pi} e^{i x P^+ z^-/2}\nonumber\\
&&\bra{\mathcal{P}^+(P')}\bar{\Psi}(0)i \sigma^{i+}\gamma_5 \Psi(z)\ket{\mathcal{P}^+(P)}\vert_{z^+=\textbf{z}_\perp=0}.
\end{eqnarray}
By inserting the initial and the final states of pion $(\pi^+(P)\ {\rm{and}}\ \pi^+(P'))$ and kaon $(K^+(P)\ {\rm{and}}\ K^+(P'))$ from Eq. (\ref{overlap}) in above equations, we obtain the quark GPDs $H(x,0,t)$ and $E_T(x,0,t)$ in overlap form of LCWFs as
\begin{widetext}
\begin{eqnarray}
H^{\pi(K)}(x,0,t)&=&\int \frac{d^2\textbf{k}_\perp}{16 \pi^3} \big[\psi_0^{\pi(K)*}(x,\textbf{k}'_\perp, \uparrow, \uparrow )\psi_0^{\pi(K)}(x,\textbf{k}_\perp, \uparrow, \uparrow)+\psi_0^{\pi(K)*}(x,\textbf{k}'_\perp, \uparrow, \downarrow) \psi_0^{\pi(K)}(x,\textbf{k}_\perp, \uparrow, \downarrow )\nonumber\\
&&+\psi_0^{\pi(K)*}(x,\textbf{k}'_\perp, \downarrow, \uparrow )\psi_0^{\pi(K)}(x,\textbf{k}_\perp, \downarrow, \uparrow )+\psi_0^{\pi(K)*}(x,\textbf{k}'_\perp, \downarrow, \downarrow )\psi_0^{\pi(K)}(x,\textbf{k}_\perp, \downarrow, \downarrow )\big],
\end{eqnarray}
\begin{eqnarray}
-\frac{i q_2}{M_\mathcal{P}}E_T^{\pi(K)}(x,0,t)&=&\int \frac{d^2\textbf{k}_\perp}{16 \pi^3} \big[\psi_0^{\pi(K)*}(x,\textbf{k}'_\perp, \uparrow, \uparrow )\psi_0^{\pi(K)}(x,\textbf{k}_\perp, \downarrow, \uparrow)+\psi_0^{\pi(K)*}(x,\textbf{k}'_\perp, \downarrow, \uparrow) \psi_0^{\pi(K)}(x,\textbf{k}_\perp, \uparrow, \uparrow )\nonumber\\
&&+\psi_0^{\pi(K)*}(x,\textbf{k}'_\perp, \uparrow, \downarrow )\psi_0^{\pi(K)}(x,\textbf{k}_\perp, \downarrow, \downarrow )+\psi_0^{\pi(K)*}(x,\textbf{k}'_\perp, \downarrow, \downarrow )\psi_0^{\pi(K)}(x,\textbf{k}_\perp, \uparrow, \downarrow )\big]
\end{eqnarray}
where the final state struck quark momentum is written as
\begin{eqnarray}
\textbf{k}'_\perp=\textbf{k}_\perp-(1-x)\textbf{q}_\perp.
\end{eqnarray}
Also, it is noticeable here that we choose the quark polarization along $y$-direction i.e. $i=2$. 
By substituting the respective wave functions for pion and kaon from Eqs. (\ref{space}), (\ref{coeff}), (\ref{bhl-pi}) and (\ref{bhl-k}), we get the explicit expressions of GPDs. For the case of pion, we have

\begin{eqnarray}
H^\pi(x,0,t)&=&\int \frac{d^2\textbf{k}_\perp}{16 \pi^3} \bigg[\big((x\mathcal{M'}^\pi+m)((1-x)\mathcal{M'}^\pi+m)-\textbf{k}'^2_\perp\big)\big((x\mathcal{M}^\pi+m)((1-x)\mathcal{M}^\pi+m)-\textbf{k}^2_\perp\big)\nonumber\\
&&+\big(\mathcal{M'}^\pi+2 m\big)\big(\mathcal{M}^\pi+2 m\big)\bigg]\frac{\varphi^{\pi*}(x,\textbf{k}'_\perp)\varphi^{\pi}(x,\textbf{k}_\perp)}{\omega'_1 \omega'_2 \omega_1 \omega_2},
\end{eqnarray}
\begin{eqnarray}
E_T^\pi(x,0,t)&=&2 M_\pi(1-x)\int \frac{d^2 {\bf k}_\perp}{16 \pi^3}\bigg[(\mathcal{M}'^\pi+2 m)\left(x(1-x)\mathcal{M}^{\pi 2}+m(\mathcal{M}^\pi+m)-{\bf k}^2_\perp\right)\bigg]
\frac{\varphi^{\pi*}(x,\textbf{k}'_\perp)\varphi^{\pi}(x,\textbf{k}_\perp)}{\omega'_1 \omega'_2 \omega_1 \omega_2},
\end{eqnarray}
with
\begin{eqnarray}
\mathcal{M}^\pi=\sqrt{\frac{m^2+\textbf{k}^2_\perp}{x(1-x)}},\ \ \ \ \
\mathcal{M'}^\pi=\sqrt{\frac{m^2+\textbf{k}'^2_\perp}{x(1-x)}},
\end{eqnarray}
in initial and final states, respectively.
Meanwhile, the explicit expressions of the kaon GPDs are given by~\cite{Meissner:2008ay}
 \begin{eqnarray}
H^K(x,0,t)&=&\int \frac{d^2\textbf{k}_\perp}{16 \pi^3} \bigg[\big((x\mathcal{M'}^K+m_1)((1-x)\mathcal{M'}^K+m_2)-\textbf{k}'^2_\perp\big)\big((x\mathcal{M}^K+m_1)((1-x)\mathcal{M}^K+m_2)-\textbf{k}^2_\perp\big)\nonumber\\
&+&\big(\mathcal{M'}^K+ m_1+m_2\big)\big(\mathcal{M}^K+m_1+m_2\big)\bigg]\frac{\varphi^{K*}(x,\textbf{k}'_\perp)\varphi^{K}(x,\textbf{k}_\perp)}{\omega'_1 \omega'_2 \omega_1 \omega_2},
\end{eqnarray}
\begin{eqnarray}
E_T^K(x,0,t) &=& 2 M_K(1-x)\int \frac{d^2 {\bf k}_\perp}{16 \pi^3}\bigg[(\mathcal{M}'^K+ m_1+m_2)\left((x\mathcal{M}^K+m_1)((1-x)\mathcal{M}^K+m_2)-{\bf k}^2_\perp\right)\bigg]\nonumber\\
&\times &\frac{\varphi^{K *}(x,\textbf{k}'_\perp)\varphi^{K}(x,\textbf{k}_\perp)}{\omega'_1 \omega'_2 \omega_1 \omega_2},
\end{eqnarray}
\end{widetext}
with
\begin{eqnarray}
\mathcal{M}^K=\sqrt{\frac{m_1^2+\textbf{k}^2_\perp}{x}+\frac{m_2^2+\textbf{k}^2_\perp}{1-x}},\nonumber\\
\mathcal{M'}^K=\sqrt{\frac{m_1^2+\textbf{k}'^2_\perp}{x}+\frac{m_2^2+\textbf{k}'^2_\perp}{1-x}},
\end{eqnarray}
in the initial and  the final states, respectively.
Here, $t=-\textbf{q}^2_\perp$ is denoted as the total momentum transferred to the meson. The detailed discussion on graphical representation of GPD in case of kaon has been already explained in Ref. \cite{kaon_gpd}.

\begin{figure*}
\centering
\begin{minipage}[c]{1\textwidth}
(a)\includegraphics[width=.45\textwidth]{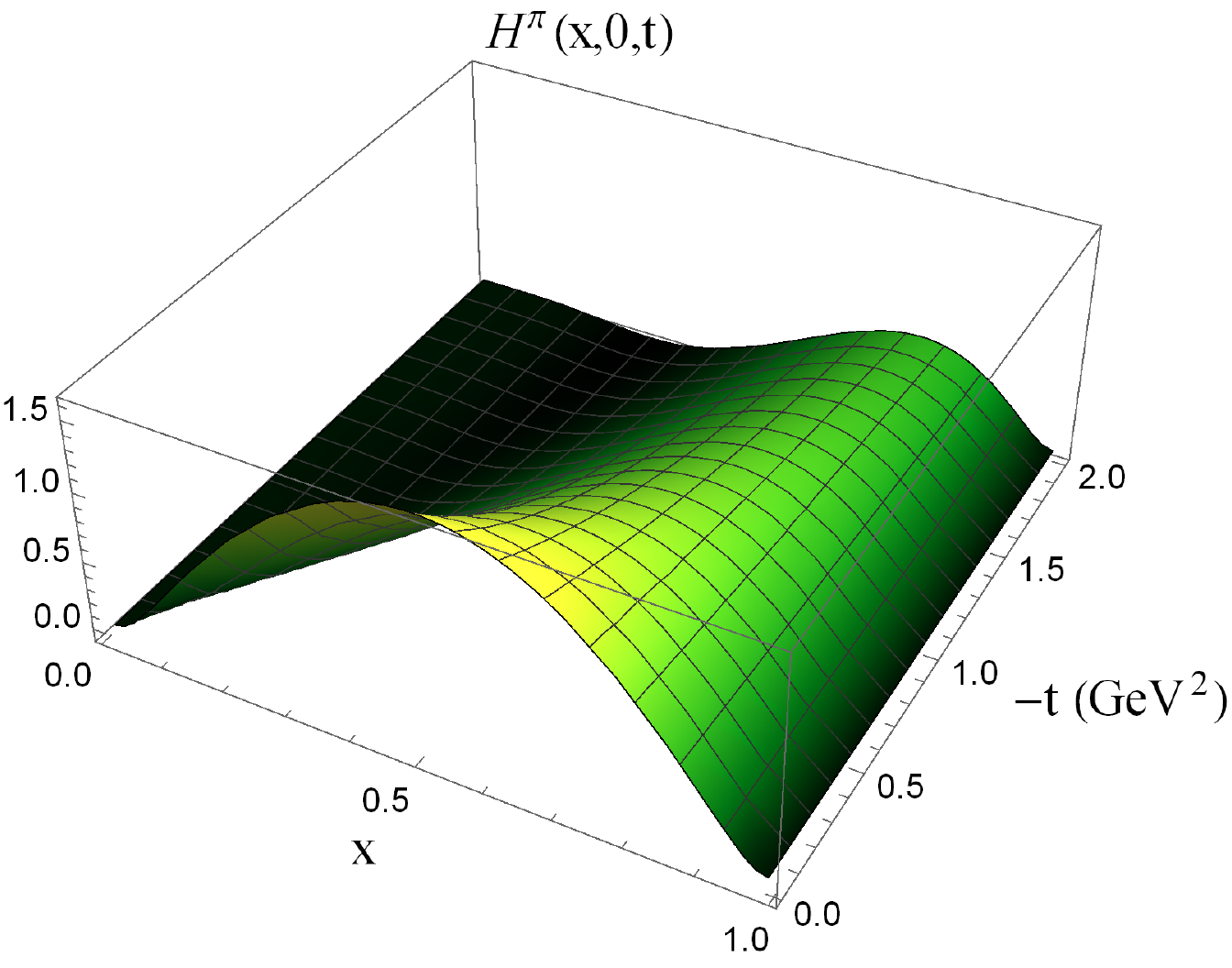}
(b)\includegraphics[width=.45\textwidth]{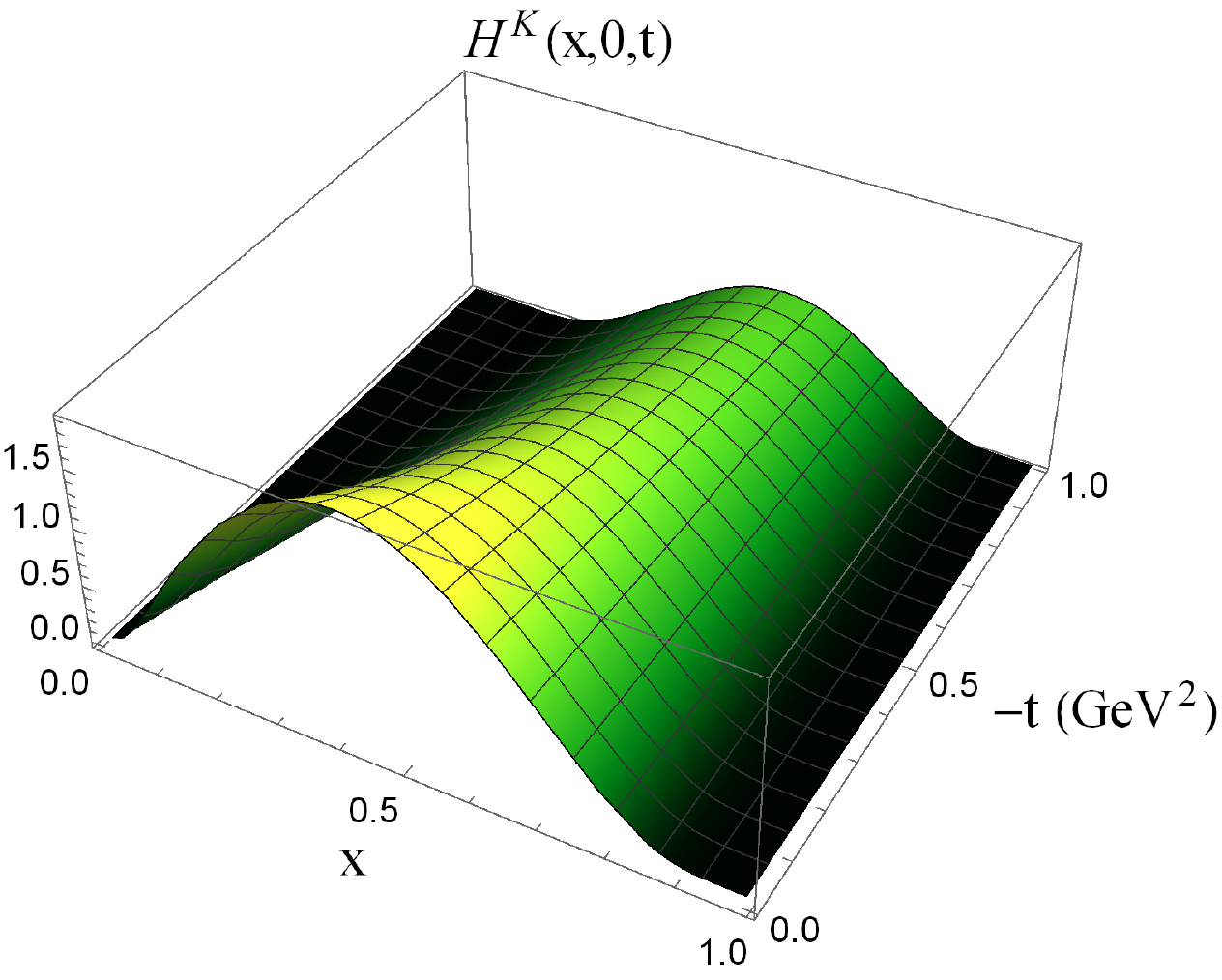}
\end{minipage}
\begin{minipage}[c]{1\textwidth}
(c)\includegraphics[width=.45\textwidth]{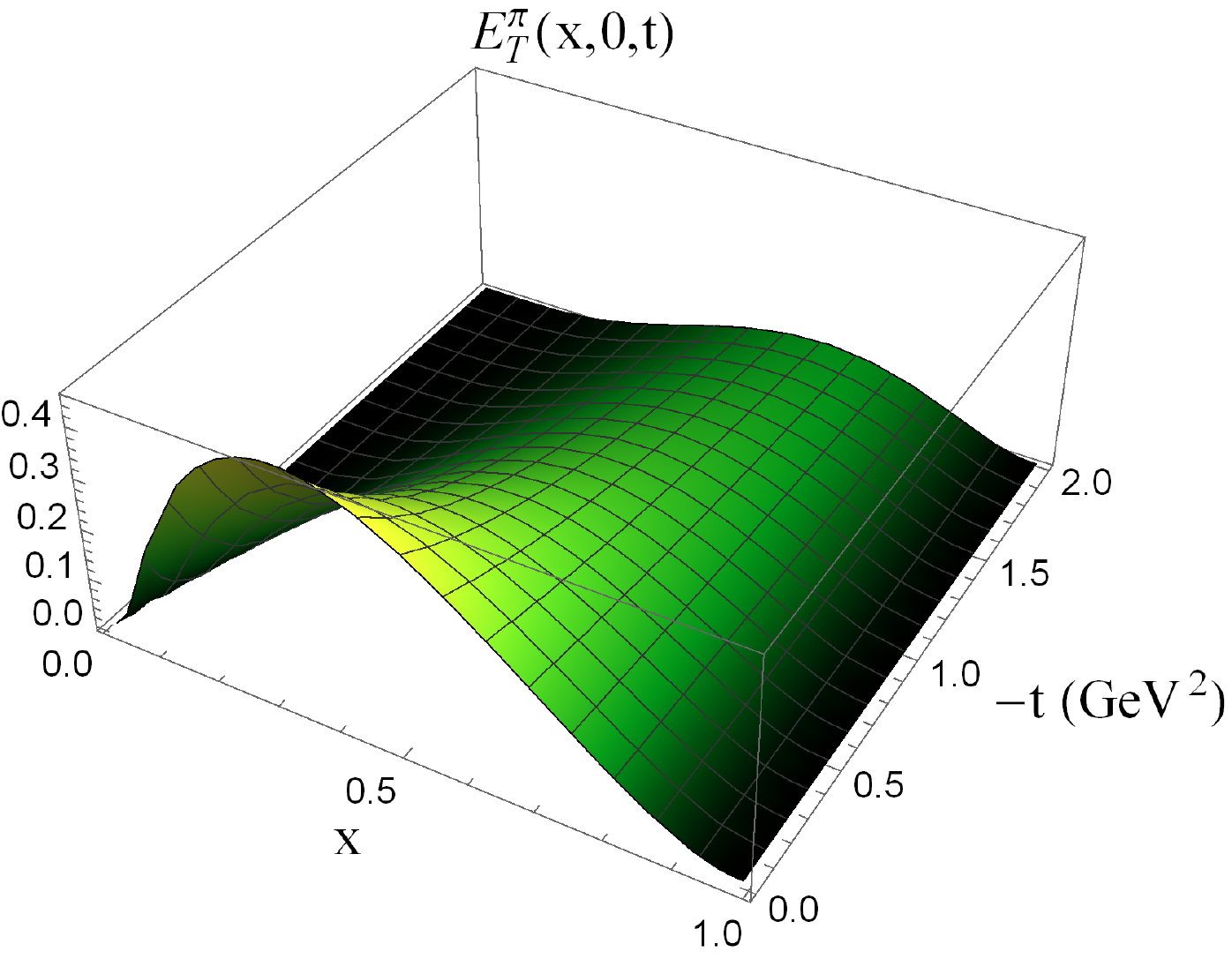}
(d)\includegraphics[width=.45\textwidth]{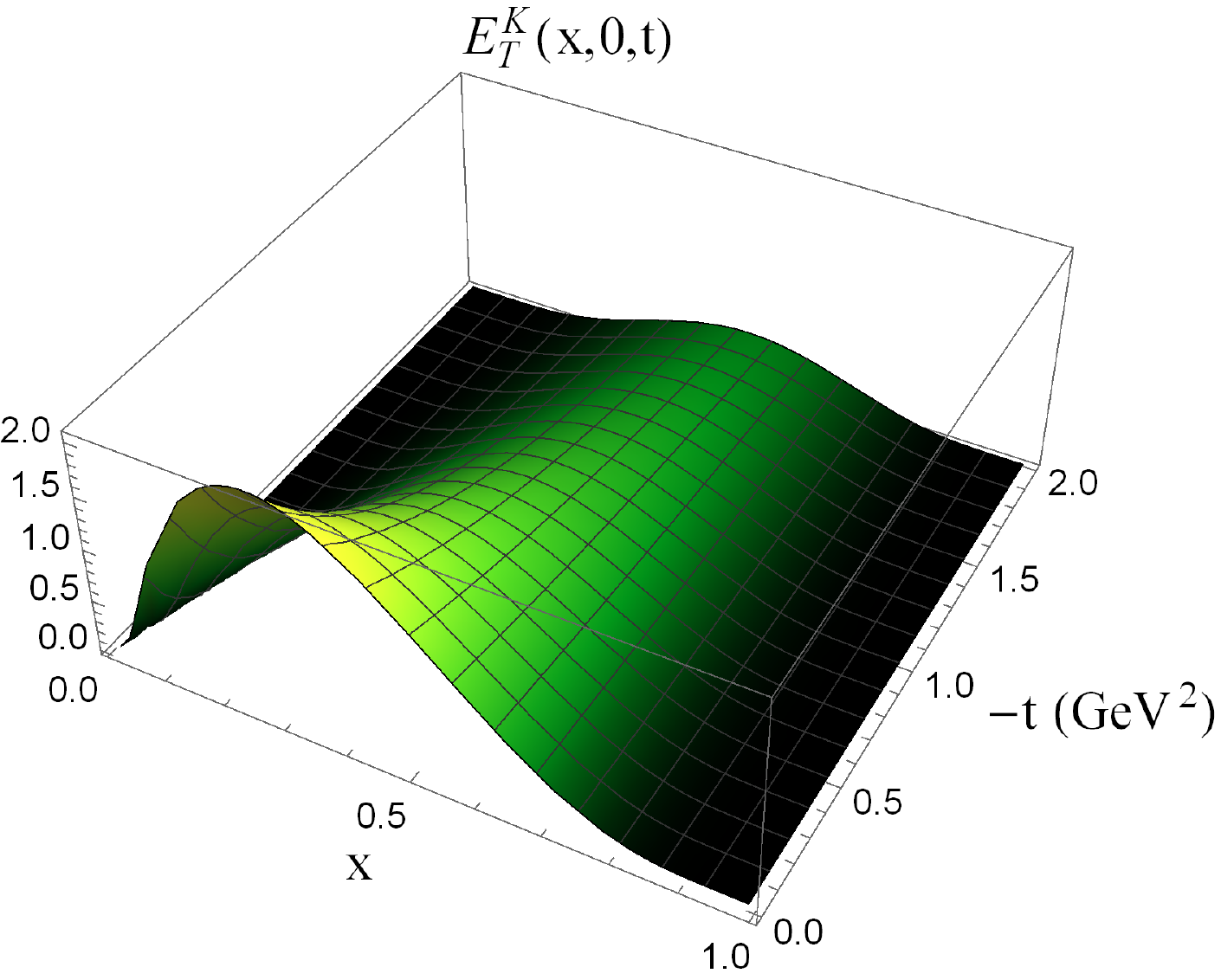}
\end{minipage}
\caption{The chiral-even $H(x,0,t)$ and the chiral-odd $E_T(x,0,t)$ GPDs as a function of $x$ and $-t$ (in $\rm GeV^2$) for pion (left panel) and for kaon (right panel).}
\label{gpd}
\end{figure*}

We use the parameters mentioned in Table \ref{parameters} to calculate the GPDs $H(x,0,t)$ and $E_T(x,0,t)$ of $u$-quark in light pseudoscalar mesons. To understand the dependence of the $u$-quark GPD on $x$ and $-t$, we illustrate the 3-D graphical representation of $H$ and  $E_T$ GPDs in Fig. \ref{gpd} for the pion (left panel) and the kaon (right panel). The unpolarized quark distribution in the pion with respect to 
the longitudinal momentum fraction $x$ is maximum at the central value ($x=0.5$) when the momentum transferred to the pion is zero. Unlike the unpolarized GPD $H$, the peak of the chiral-odd GPD $E_T$ in pion appears at below the central value of $x$ when the momentum transfer is zero. As the value of momentum transferred $-t$ increases, the peak shifts towards higher values of $x$ and the magnitude of distribution becomes lower.
Unlike pion GPD $H$, the kaon case has the maximum at lower $x$ ($<0.5$) when $t=0$. This is because of the presence of strange quark having larger mass. While with increasing $-t$, the peaks along $x$ get shifted to larger values of $x$ same as pion unpolarized GPD. This is a model independent behavior of GPDs which has been observed in other phenomenological models for pion~\cite{non-zero-skewness} as well as for nucleon~\cite{Chakrabarti:2013gra,Mondal:2015uha,Chakrabarti:2015ama,Maji:2017ill}. We also observe that similar to the unpolarized quark GPD $H$, the transversely-polarized quark GPD $E_T$ is broader in pion than that in kaon.  At large $x$ the kaon GPDs fall faster compare to the GPDs in pion.\\


\section{VI Transverse momentum-dependent parton distributions (TMDs)}
\begin{figure*}
\centering
\begin{minipage}[c]{1\textwidth}
(a)\includegraphics[width=.45\textwidth]{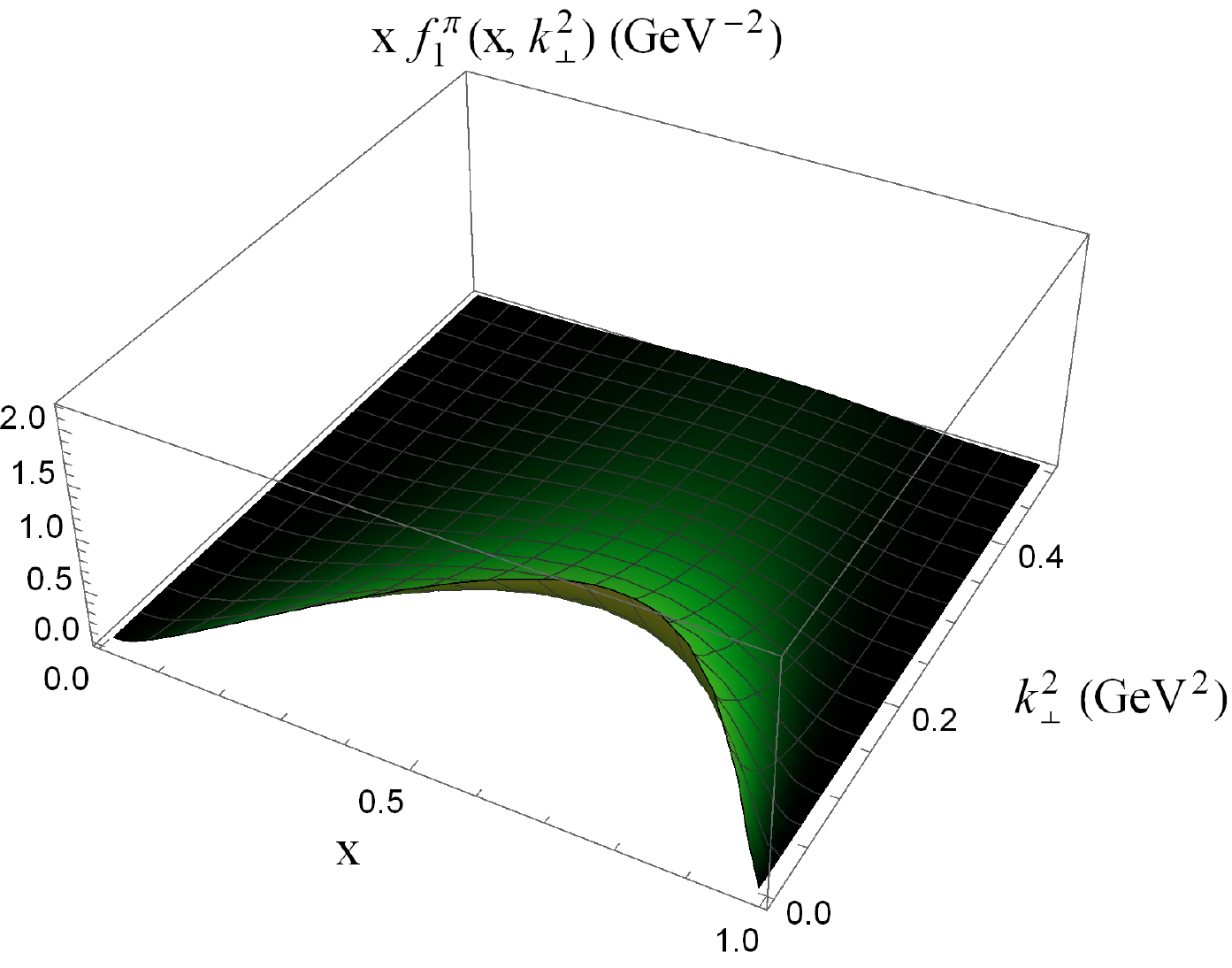}
(b)\includegraphics[width=.45\textwidth]{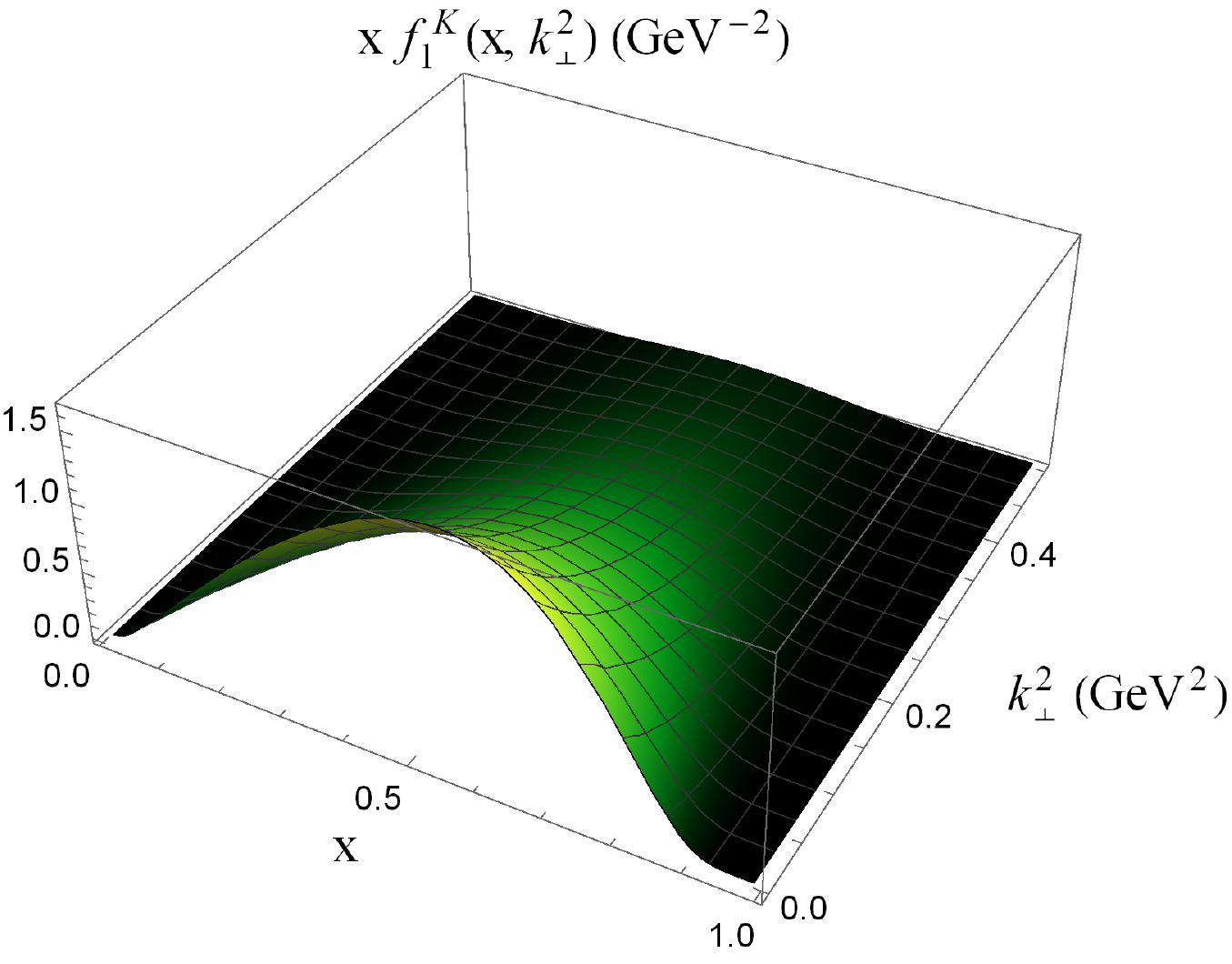}
\end{minipage}
\begin{minipage}[c]{1\textwidth}
(c)\includegraphics[width=.45\textwidth]{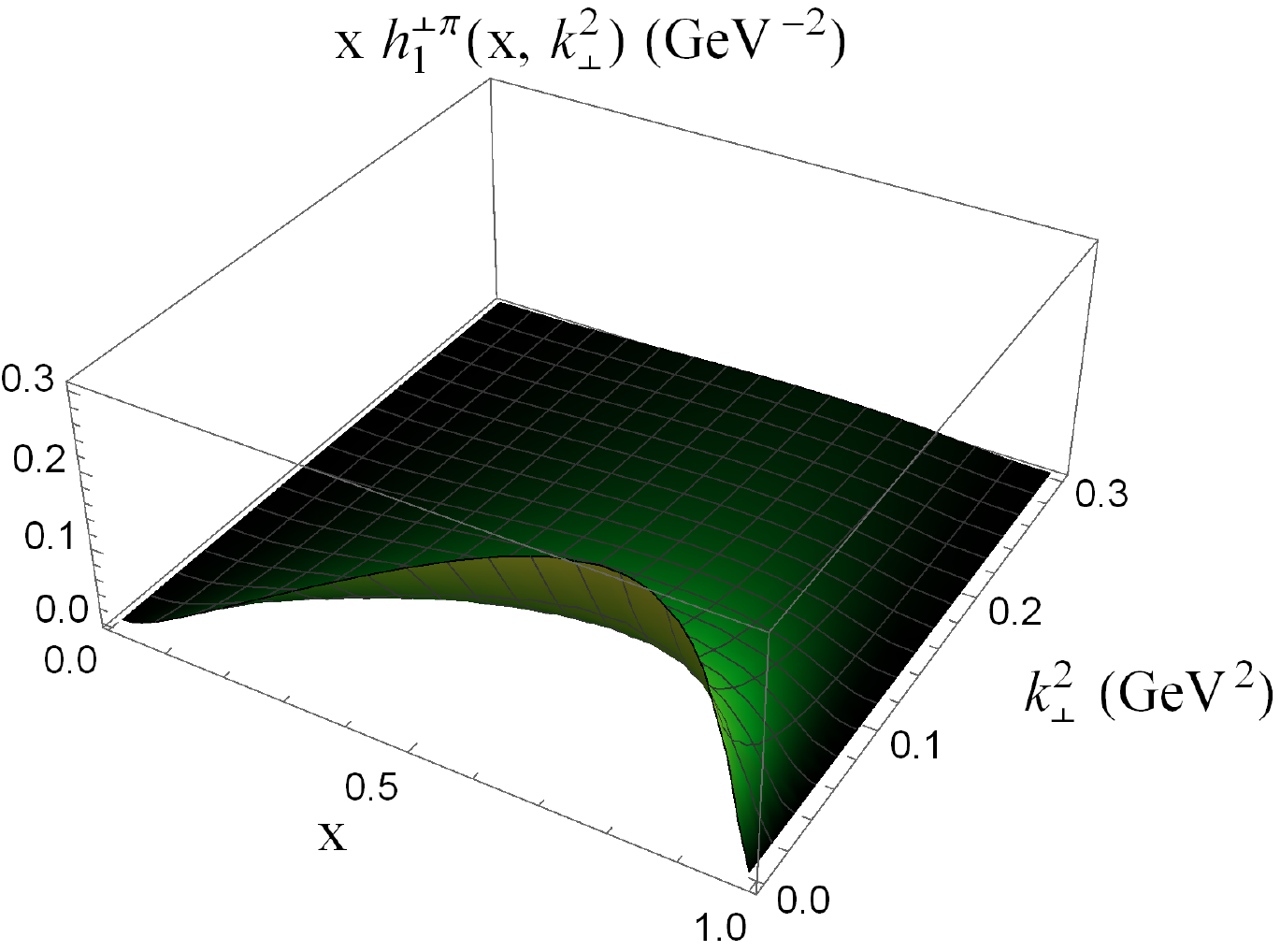}
(d)\includegraphics[width=.45\textwidth]{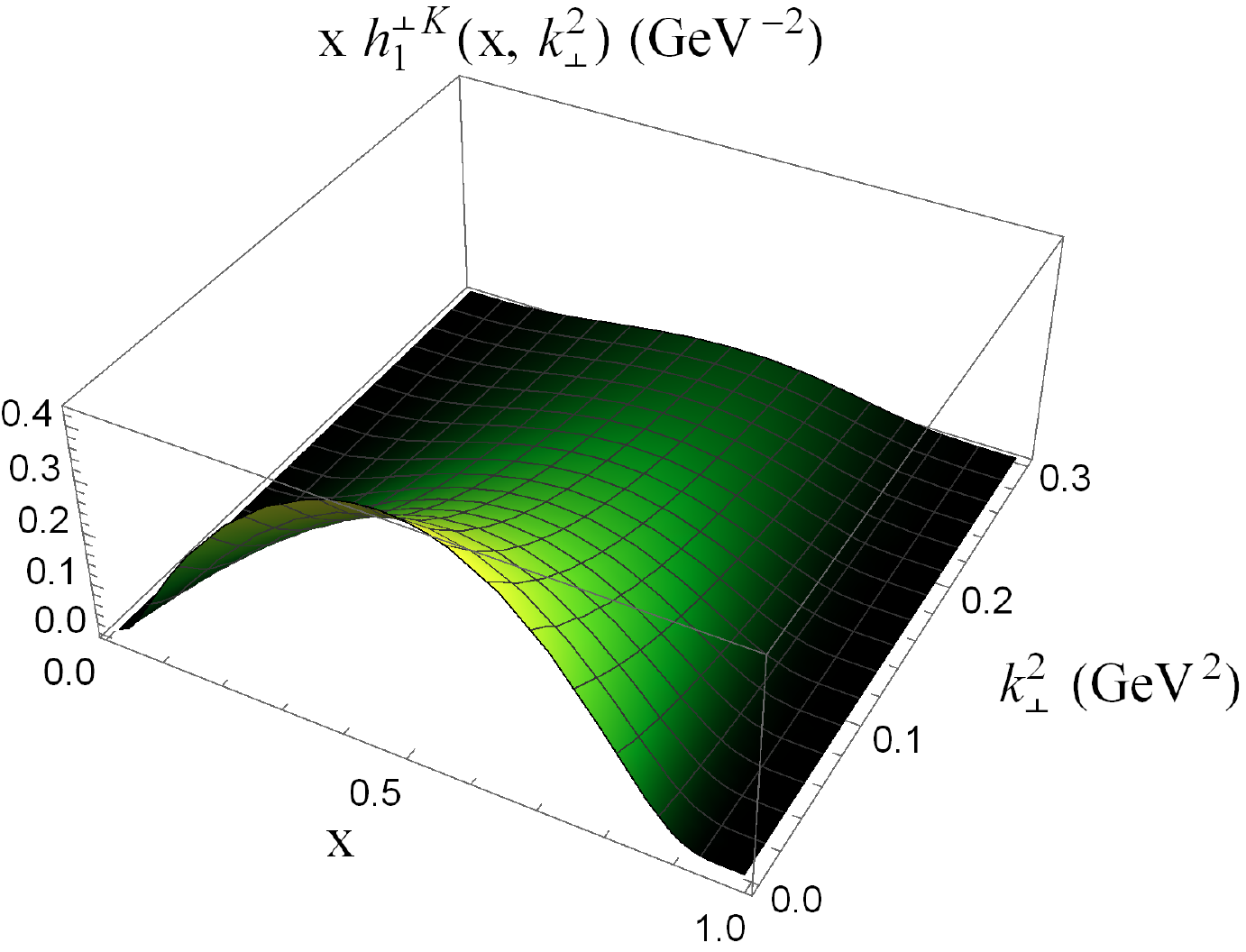}
\end{minipage}
\caption{The unpolarized and the Boer-Mulders TMDs multiplied by $x$ i.e. $ x f_1(x,\textbf{k}^2_\perp)$ and $ x h_1^{\perp\pi}(x,\textbf{k}^2_\perp)$ respectively, as a function of $x$ and $\textbf{k}^2_\perp$ (in $\rm GeV^2$) for pion (left panel) and for kaon (right panel).}
\label{3d-tmd}
\end{figure*}

TMDs provide the distribution of partons in momentum space and are functions of longitudinal momentum fraction $x=k^+/P^+$ and transverse momentum $\textbf{k}_\perp$ carried by the struck quark. To evaluate the pion and the kaon TMDs, the unintegrated quark-quark correlator can be defined as \cite{tmd-def, tmd-def1}
\begin{widetext}
\begin{eqnarray}
\Phi^{[\Gamma]\mathcal{P}}(x,\textbf{k}_\perp;S)=\frac{1}{2}\int \frac{dz^-}{2\pi} \frac{d^2 \textbf{z}_\perp}{(2\pi)^2} e^{i k.z/2}\bra{\mathcal{P}^+(P),S}\bar{\Psi}(0)\Gamma \mathcal{L}^{\dagger}(\mathbf{0}|n) \mathcal{L}\Psi(z)\ket{\mathcal{P}^+(P),S}\vert_{z^+=0}.
\end{eqnarray}
where in the light-front gauge, $A^+=0$ the gauge link is given by
 \begin{eqnarray}
 	&&\mathcal{L}_{A^+=0}(\mathbf{z}_\perp|n)= \mathcal{P} \exp\left(-ig \int_{\mathbf{z}_\perp}^{\infty} \mathrm{d} \mathbf{\eta}_\perp \cdot \mathbf{A}_\perp(\eta^-=n \cdot \infty,\mathbf{z}_{\perp})\right)
 \label{gauge-link}
 \end{eqnarray}
At the leading twist there are two independent TMDs for pseudoscalar mesons. The unpolarized quark TMD, $f_1(x, k_\perp)$, describes the momentum distribution of unpolarized quarks within the meson while the polarized quark TMD, $h_1^{\perp}(x, k_\perp)$ (the Boer-Mulders TMD) describes the spin-orbit correlations of transversely polarized quarks within the pion.
%
The unpolarized and the Boer-Mulders TMDs are expressed as \cite{tmd-def2, tmd-def3}
\begin{eqnarray}
f^{\mathcal{P}}_1(x,{\bf k}_\perp)&=&{\rm Tr}(\Phi^{[\gamma^+]}),\\
h_1^{\perp \mathcal{P}}(x,{\bf k}_\perp)&=&\frac{\epsilon^{ij}{k}_\perp^j M_\mathcal{P}}{ {\bf k}^2_\perp} {\rm Tr}(\Phi^{[i \sigma^{i+}\gamma_5]}).
\end{eqnarray}

By taking the gauge link unity and $\Gamma=\gamma^+$, we get the explicit expressions of unpolarized pion TMD $f_1^\pi(x,\textbf{k}_\perp^2)$ and unpolarized kaon TMD $f_1^K(x,\textbf{k}_\perp^2)$ using the states of respective mesons. The overlap form of unpolarized TMD $f_1(x,\textbf{k}_\perp^2)$ reads
\begin{eqnarray}
f_1^{\pi(K)}(x,\textbf{k}^2_\perp)&=&\frac{1}{16 \pi^3} \big[\mid{ \psi_0^{\pi(K)}(x,\textbf{k}_\perp, \uparrow, \uparrow )}\mid^2 + \mid \psi_0^{\pi(K)}(x,\textbf{k}_\perp, \uparrow, \downarrow)\mid^2 + \mid \psi_0^{\pi(K)} (x,\textbf{k}_\perp, \downarrow, \uparrow )\mid^2 \nonumber\\
&&+\mid \psi_0^{\pi(K)}(x,\textbf{k}_\perp, \downarrow, \downarrow )\mid^2 \big].
\end{eqnarray}
 On the other hand, to generate the non-zero Boer-Mulders function, one needs to take into account the gauge link. Physically, this is equivalent to taking into account the initial or the final state interactions of the active quark with the target remnant. This has been referred collectively as gluon rescattering karnel $G (x, \mathbf{k}- \mathbf{k}^\prime)$~\cite{tmd-def,Ahmady:2019yvo,Bacchetta:2008af} and one defines the Boer-Mulders function in such a way that
\begin{eqnarray}
{\bf k}^2_\perp h_1^\perp(x,{\bf k}^2_\perp)&=&M_{\mathcal{P}}\int \frac{d^2{\bf q}_\perp}{16 \pi^3} i G(x,{\bf q}_\perp)\Big[k_R \Big( \psi_0^{*\pi(K)}(x,\textbf{k}'_\perp, \downarrow, \uparrow)\psi_0^{\pi(K)}(x,\textbf{k}'_\perp, \uparrow, \uparrow)\nonumber\\
&&+\psi_0^{*\pi(K)}(x,\textbf{k}'_\perp, \downarrow, \downarrow)\psi_0^{\pi(K)}(x,\textbf{k}_\perp, \uparrow, \downarrow)\Big)-k_L\Big( \psi_0^{*\pi(K)}(x,\textbf{k}'_\perp, \uparrow, \uparrow)\psi_0^{\pi(K)}(x,\textbf{k}'_\perp, \downarrow, \uparrow)\nonumber\\
&&+\psi_0^{*\pi(K)}(x,\textbf{k}'_\perp, \uparrow, \downarrow)\psi_0^{\pi(K)}(x,\textbf{k}_\perp, \downarrow, \downarrow)\Big)\Big],
\end{eqnarray}
where perturbative Abelian gluon rescattering kernel is given by ~\cite{Bacchetta:2008af,Wang:2017onm}
\begin{eqnarray}
iG(x,{\bf q}_\perp)=\frac{c_F \alpha_s}{2 \pi}\frac{1}{{\bf q}^2_\perp},
\end{eqnarray}
with ${\bf q}_\perp={\bf k}_\perp-{\bf k}'_\perp$ and $\alpha_s$ is the fixed coupling constant. Using the LCWFs for the pion and kaon given in Eqs. (\ref{space}), (\ref{coeff}), (\ref{bhl-pi}) and (\ref{bhl-k}), the explicit expressions for the pion TMDs read 
\begin{eqnarray}
f_1^\pi(x,\textbf{k}^2_\perp)&=&\frac{1}{16 \pi^3} \bigg[\big((x\mathcal{M}^\pi+m)((1-x)\mathcal{M}^\pi+m)-\textbf{k}^2_\perp\big)^2
+\big(\mathcal{M}^\pi+2 m\big)^2\bigg]\frac{\mid \varphi^{\pi}(x,\textbf{k}_\perp)\mid^2}{\omega^2_1 \omega^2_2},
\end{eqnarray}
\begin{eqnarray}
h_1^{\perp\pi}(x,\textbf{k}^2_\perp)&=& \frac{M_\pi}{{\bf k}^2_\perp }\frac{c_F\alpha_s}{2 \pi}\int \frac{d^2{\bf q}_\perp}{16 \pi^3}\frac{1}{{\bf q}_\perp^2}\bigg[{\bf k}^2_\perp \left(\left(x(1-x)\mathcal{M}'^{ \pi 2}+m(\mathcal{M}'^\pi+m)\right)-{\bf k}'^2_\perp\right)(\mathcal{M}^\pi+2 m)\nonumber\\
&& - ({\bf k}^2_\perp-{\bf k}_\perp \cdot {\bf q}_\perp)(\mathcal{M}'^\pi+2 m)\left(x(1-x)\mathcal{M}^{ \pi 2}+m(\mathcal{M}^\pi+m)-{\bf k}^2_\perp\right) \bigg]\frac{ \varphi^{\pi *}(x,\textbf{k}'_\perp)\varphi^{\pi}(x,\textbf{k}_\perp)}{\omega'_1 \omega'_2\omega_1 \omega_2},
\end{eqnarray}
while, for kaon we have

\begin{eqnarray}
f^K_1(x,\textbf{k}^2_\perp)&=&\frac{1}{16 \pi^3} \bigg[\big((x\mathcal{M}^K+m_1)((1-x)\mathcal{M}^K+m_2)-\textbf{k}^2_\perp\big)^2
+\big(\mathcal{M}^K+ m_1+m_2\big)^2\bigg]  \frac{\mid \varphi^{K}(x,\textbf{k}_\perp)\mid^2}{\omega^2_1 \omega^2_2},
\end{eqnarray}
\begin{eqnarray}
h_1^{\perp K}(x,\textbf{k}^2_\perp)&=& \frac{M_K}{{\bf k}^2_\perp}\frac{c_F\alpha_s}{2 \pi}\int \frac{d^2{\bf q}_\perp}{16 \pi^3}\frac{1}{{\bf q}_\perp^2}\bigg[{\bf k}^2_\perp \left((x \mathcal{M}'^K+m_1)((1-x) \mathcal{M}'^K+m_2)-{\bf k}'^2_\perp\right)(\mathcal{M}^K+ m_1+m_2)\nonumber\\
&& - ({\bf k}^2_\perp-{\bf k}_\perp \cdot {\bf q}_\perp)(\mathcal{M}'^K+ m_1 + m_2)\left((x \mathcal{M}^K+m_1)((1-x) \mathcal{M}^K+m_2)-{\bf k}^2_\perp\right) \bigg]\nonumber\\
&&\times\frac{ \varphi^{K *}(x,\textbf{k}'_\perp)\varphi^{K}(x,\textbf{k}_\perp)}{\omega'_1 \omega'_2\omega_1 \omega_2}.
\end{eqnarray}

\end{widetext}
In Figs. \ref{3d-tmd}(a) and \ref{3d-tmd}(b), to get the combined information, we show the 3-dimensional picture of distribution $x f_1$ of a unpolarized quark in the unpolarized pion and kaon with respect to the longitudinal momentum fraction and squared of the quark transverse momentum respectively. The probability of finding the quark in pion is more as compared to kaon, if the momentum fraction carried by that quark is higher in the longitudinal direction. As we increase the $\textbf{k}_\perp^2$, the momentum distribution starts lowering down in both cases.
 The distribution decreases when the transverse momentum carried by quark increases. The probability to find the quark in pion and kaon starts decreasing and eventually becomes zero by the increase in the quark transverse momentum.
We illustrate the Boer-Mulders function generated by the perturbative rescattering kernel with $\alpha_s=0.3$ in Fig. \ref{3d-tmd}(c) and \ref{3d-tmd}(d) for the pion and the kaon, respectively. In the perturbative limit, the coupling is weak, however, there is no concord in the literature on what value of $\alpha_s$ should be taken in perturbative kernel. For example, while $\alpha_s=0.3$ has been used in Ref. \cite{Lu:2004hu}, much larger values of $\alpha_s$: $\alpha_s=1.2$ and $\alpha_s=0.911$ have been preferred to use in Ref. \cite{tmd-def} and  in Ref. \cite{Wang:2017onm}, respectively. The use of such large values of $\alpha_s$ contradicts the weak coupling hypothesis leading to perturbative kernel. On the other hand, taking $\alpha_s \sim 1$ may perhaps be considered as a phenomenological way to account for non-perturbative effects to some extent. In this model, we observe that the Boer-Mulders TMD exhibits a similar behavior as unpolarized TMD for both the pion and the kaon.

\begin{figure*}
\centering
\begin{minipage}[c]{0.98\textwidth}
(a)\includegraphics[width=5.2cm]{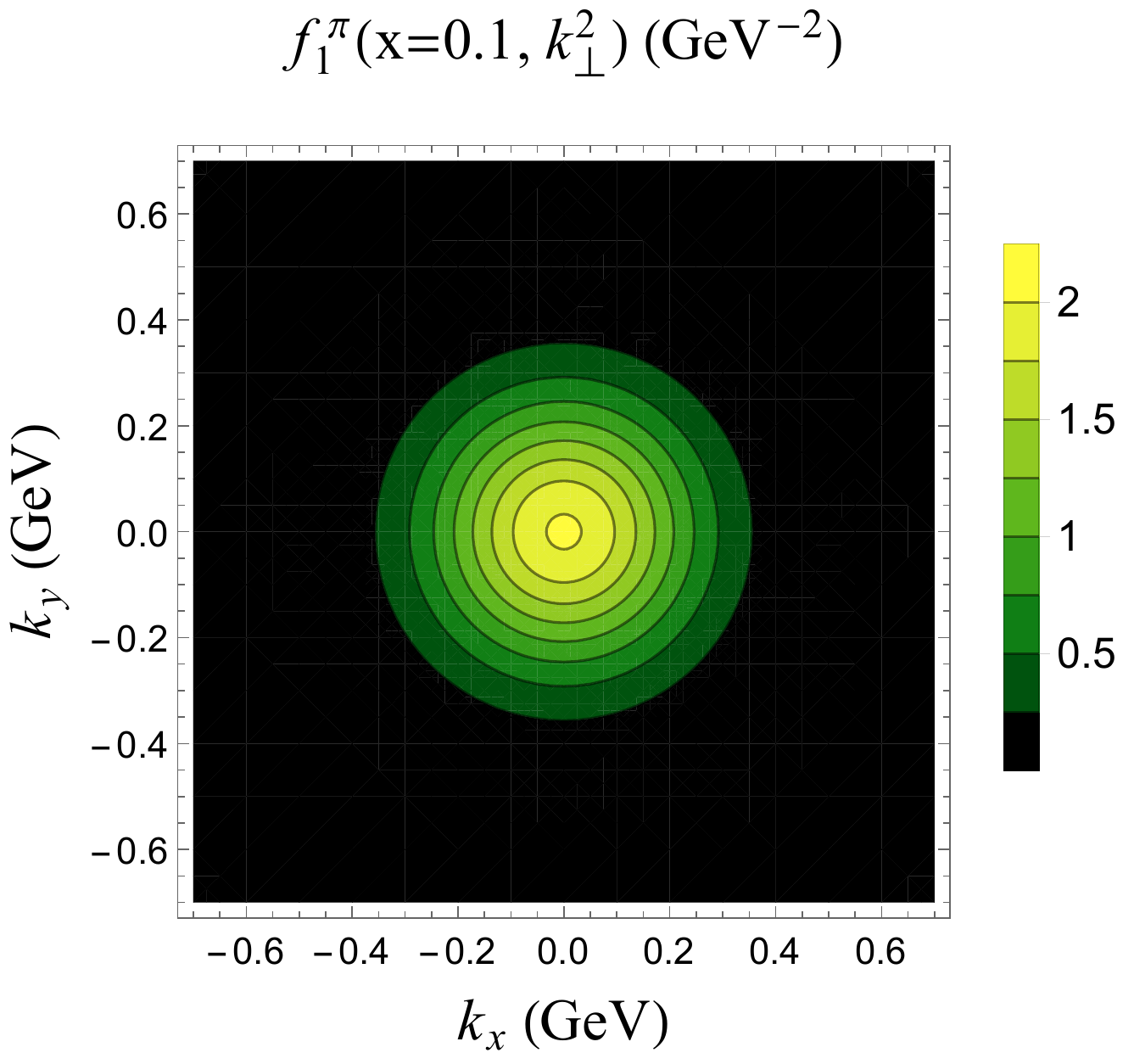}
\hspace{0.05cm}
(b)\includegraphics[width=5.2cm]{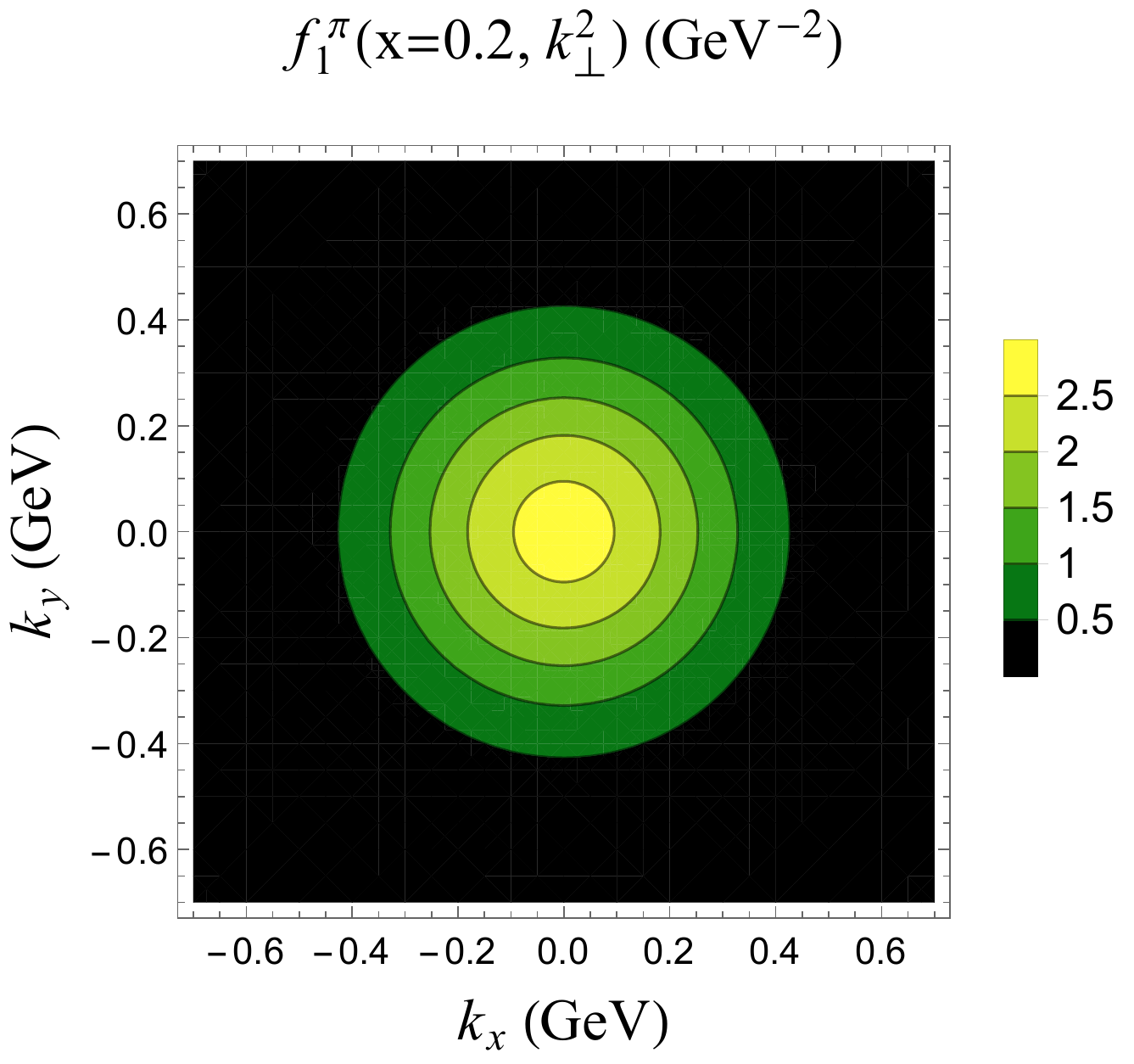}
\hspace{0.05cm}
(c)\includegraphics[width=5.2cm]{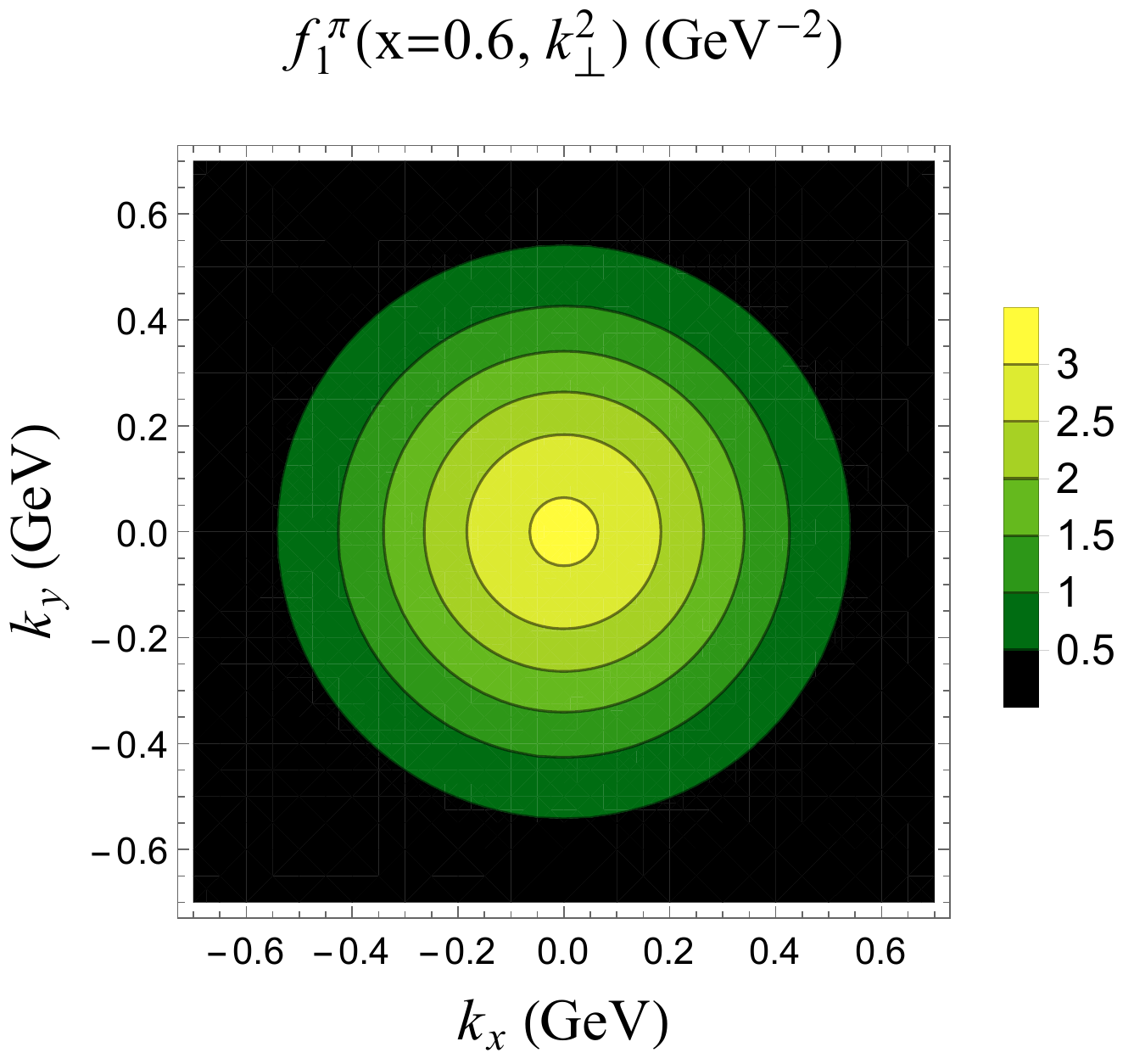}
\hspace{0.05cm}
\end{minipage}
\begin{minipage}[c]{0.98\textwidth}
(d)\includegraphics[width=5.2cm]{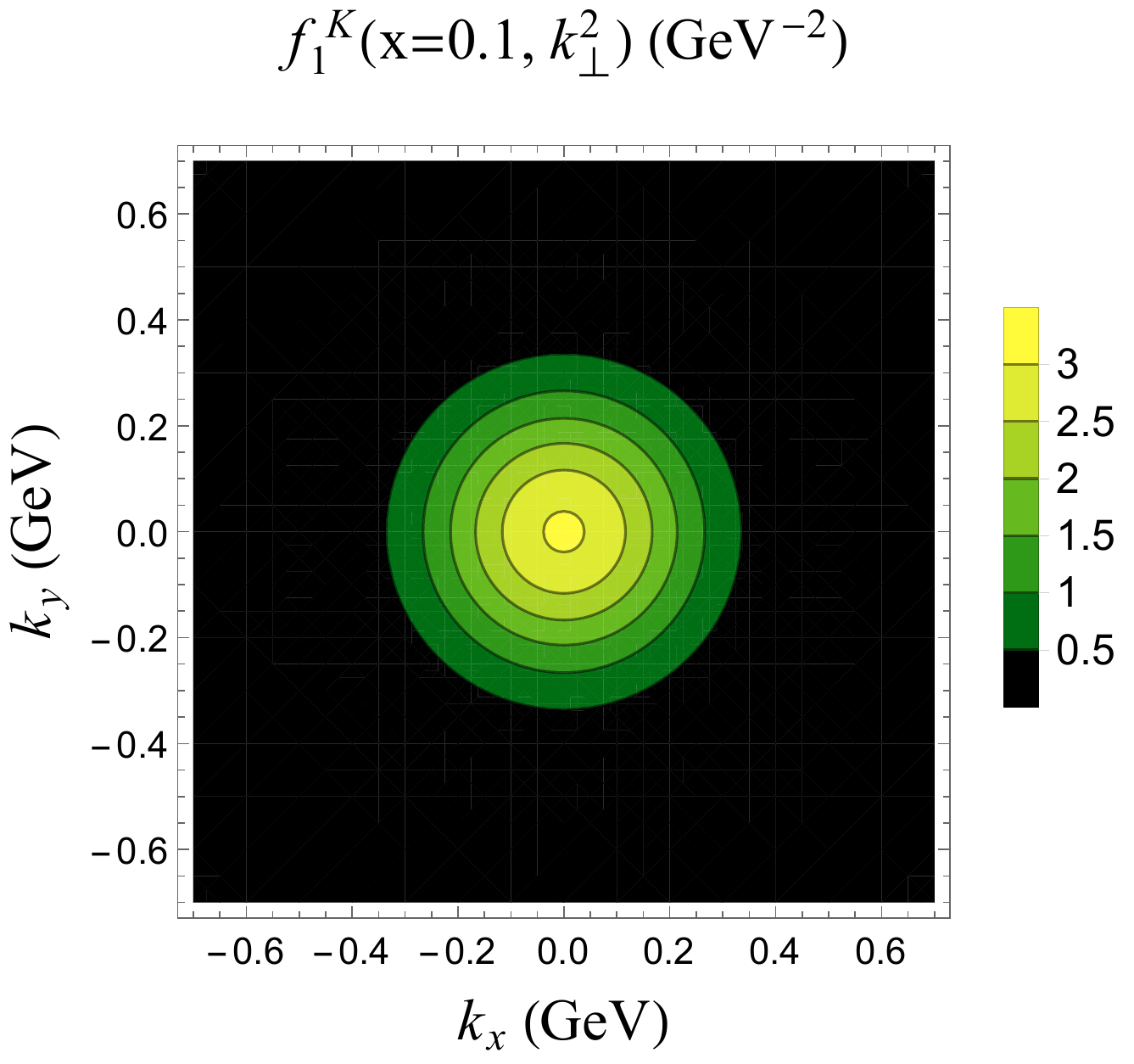}
\hspace{0.05cm}
(e)\includegraphics[width=5.2cm]{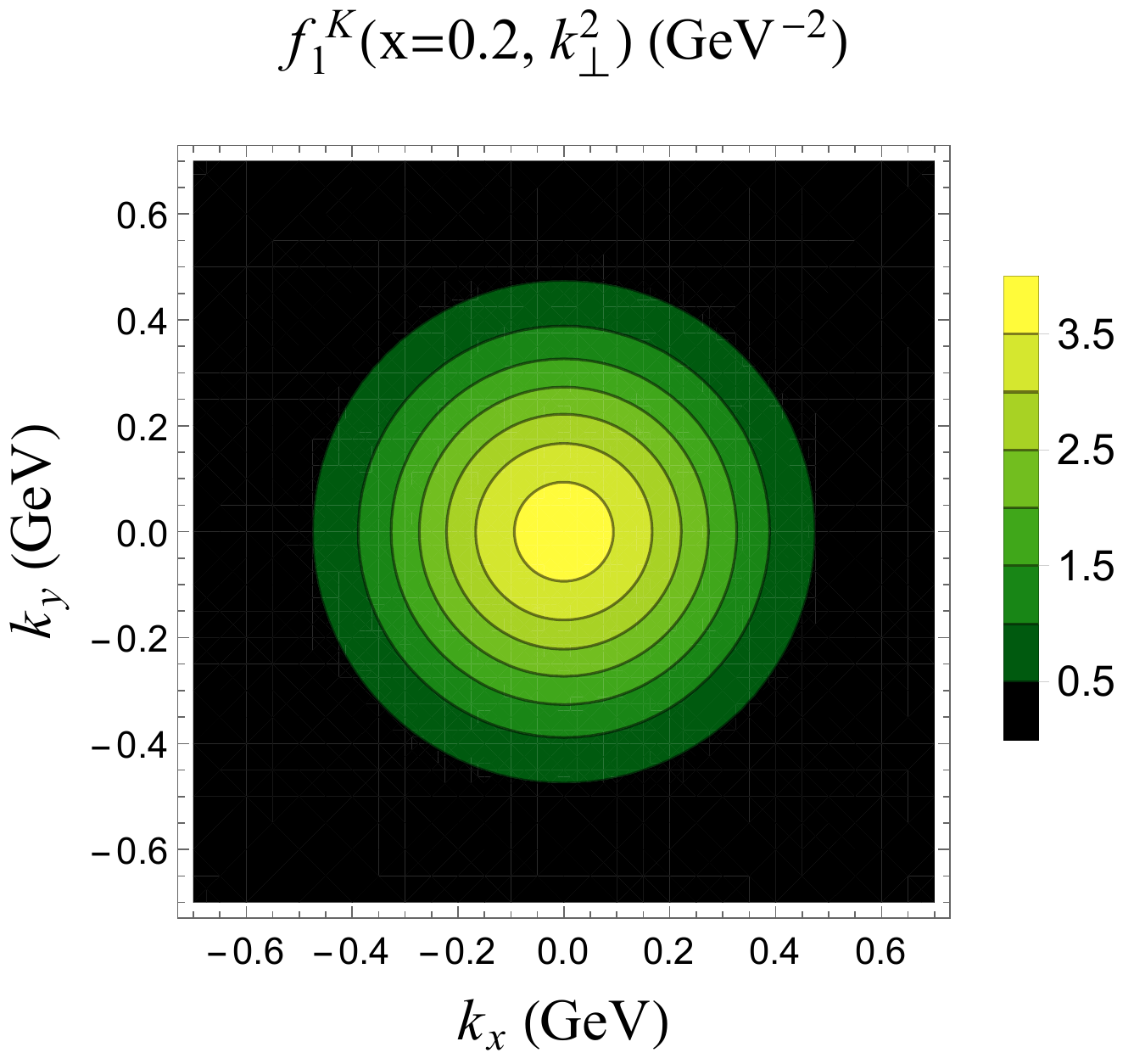}
\hspace{0.05cm}
(f)\includegraphics[width=5.2cm]{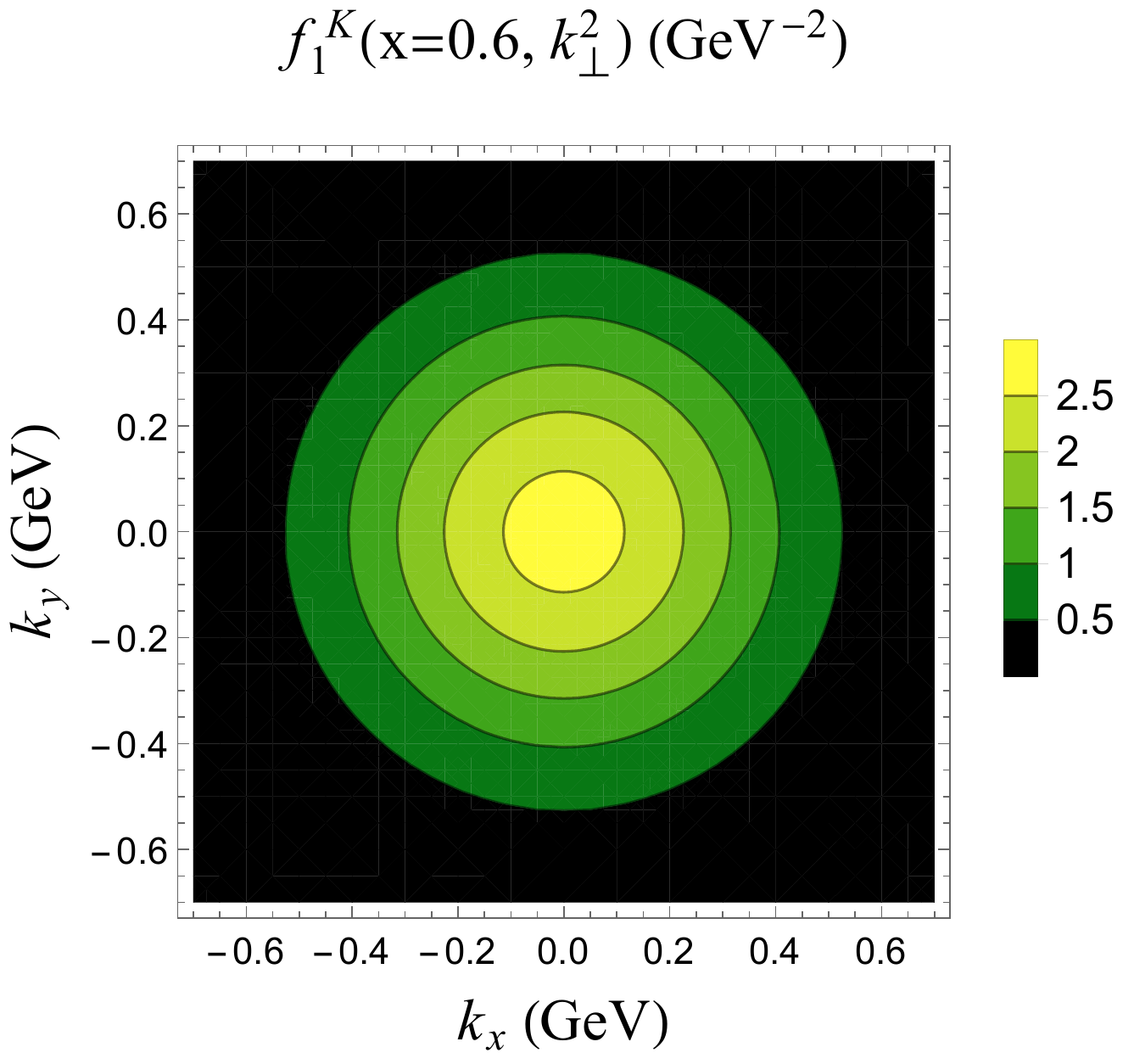}
\hspace{0.05cm}
\end{minipage}
\caption{The unpolarized TMD $ f_1(x,\textbf{k}^2_\perp)$  for pion (upper panel) and kaon (lower panel), at different values of $x$ : $x=0.1$ (left panel), $x=0.2$ (middle panel) and $x=0.6$ (right panel) presented in transverse momentum plane $(k_x, k_y)$.}
\label{cp-pi-tmd}
\end{figure*}

\begin{figure*}
\centering
\begin{minipage}[c]{0.98\textwidth}
(a)\includegraphics[width=5.2cm]{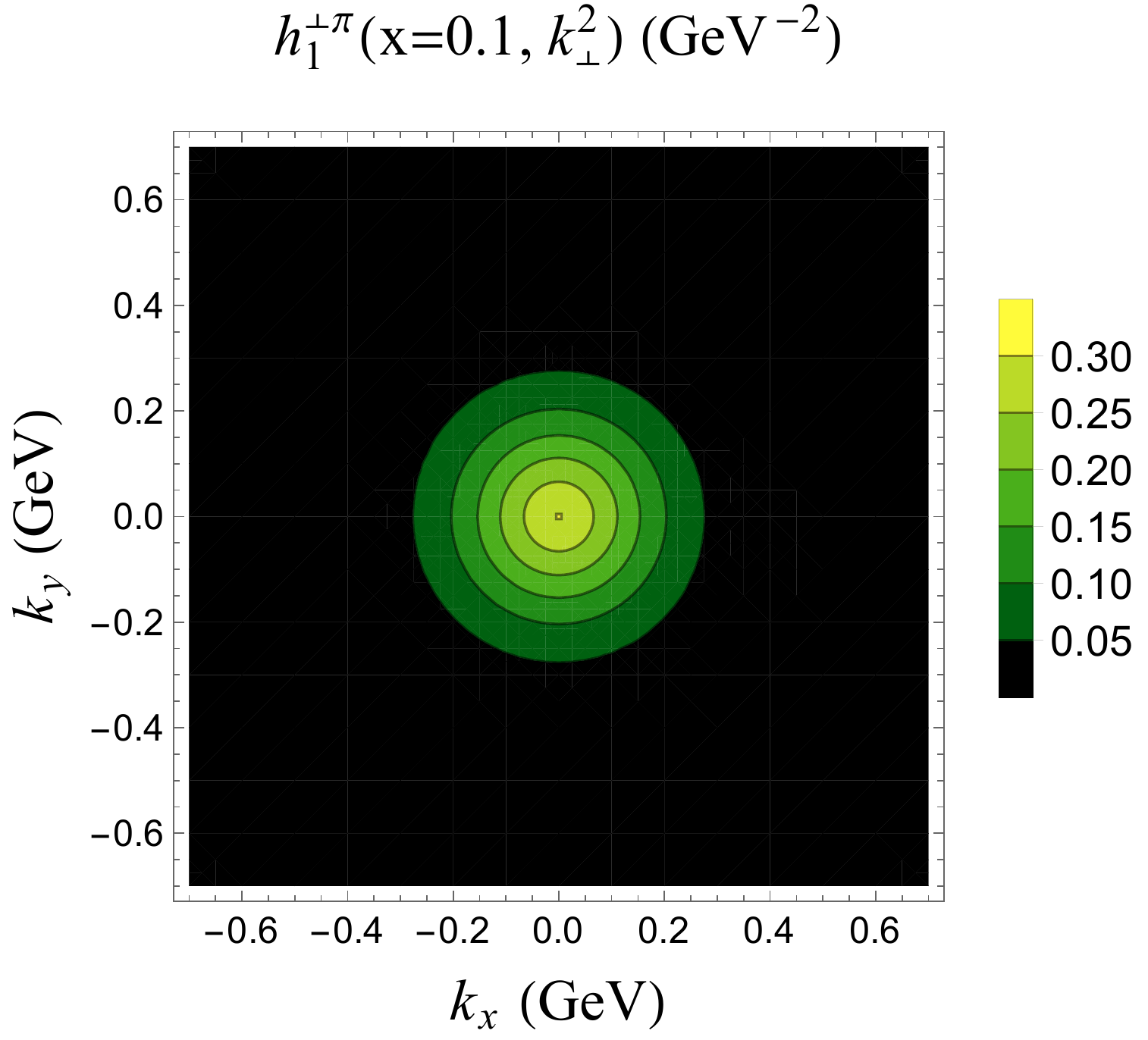}
\hspace{0.05cm}
(b)\includegraphics[width=5.2cm]{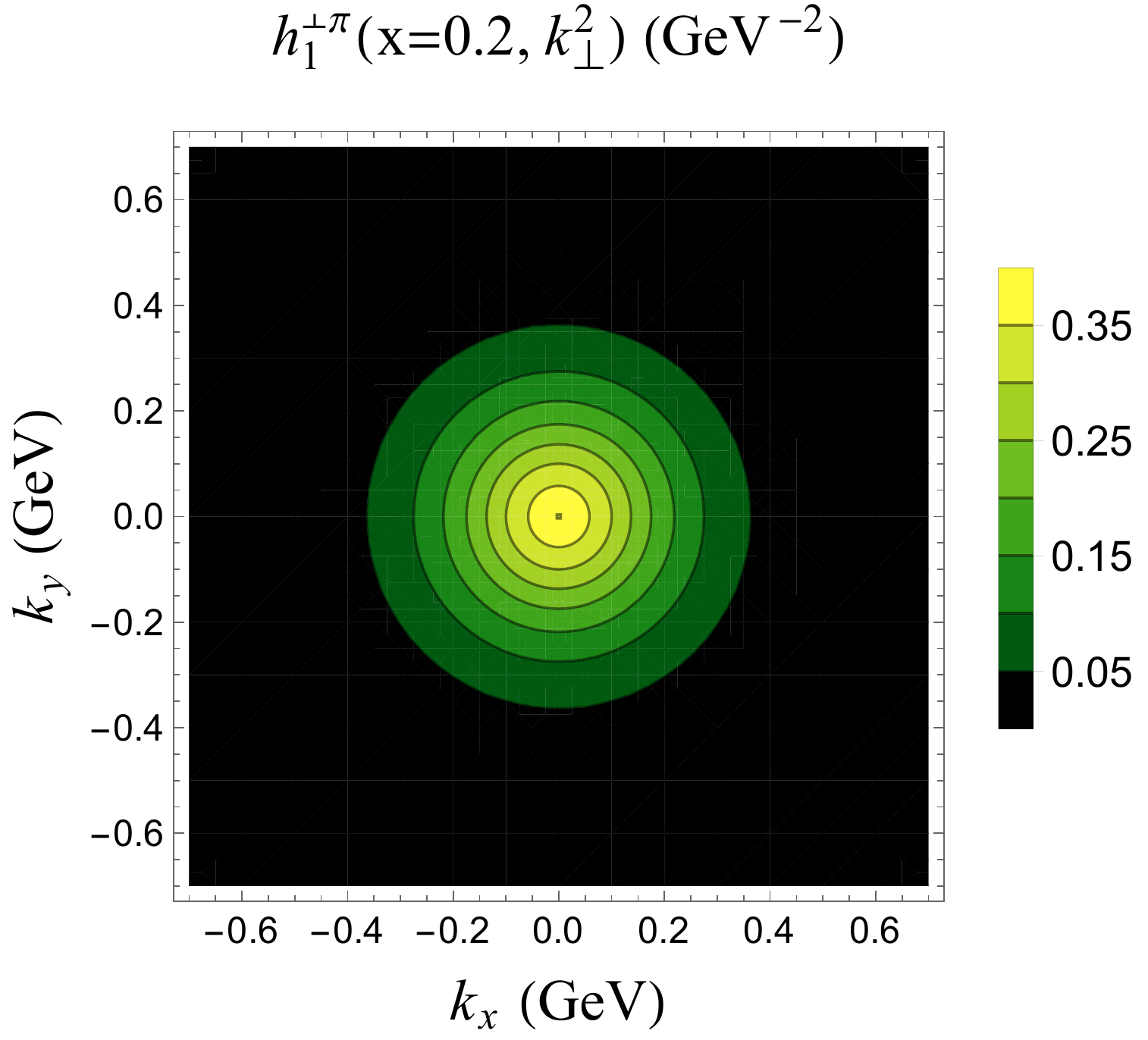}
\hspace{0.05cm}
(c)\includegraphics[width=5.2cm]{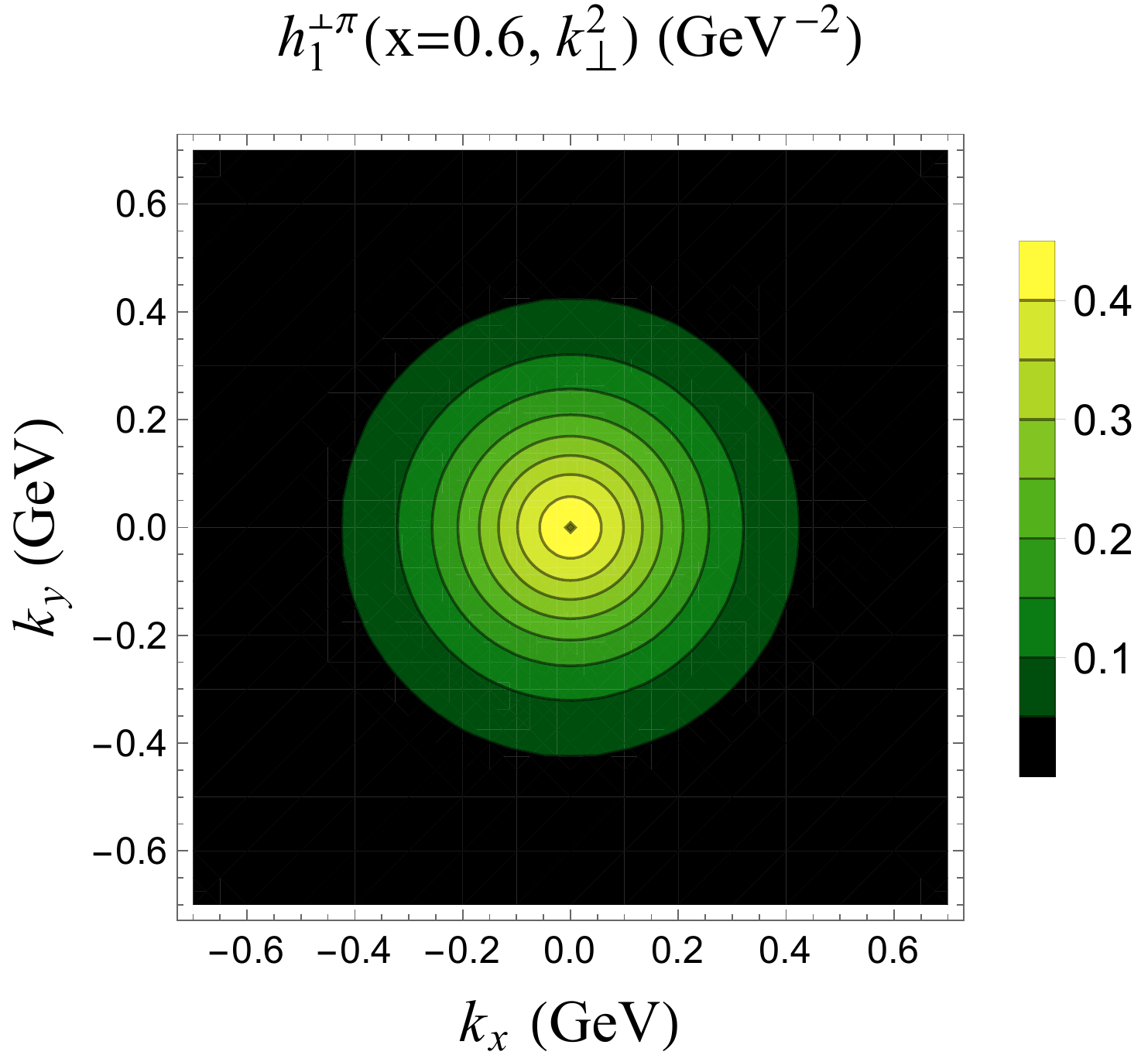}
\hspace{0.05cm}
\end{minipage}
\begin{minipage}[c]{0.98\textwidth}
(d)\includegraphics[width=5.2cm]{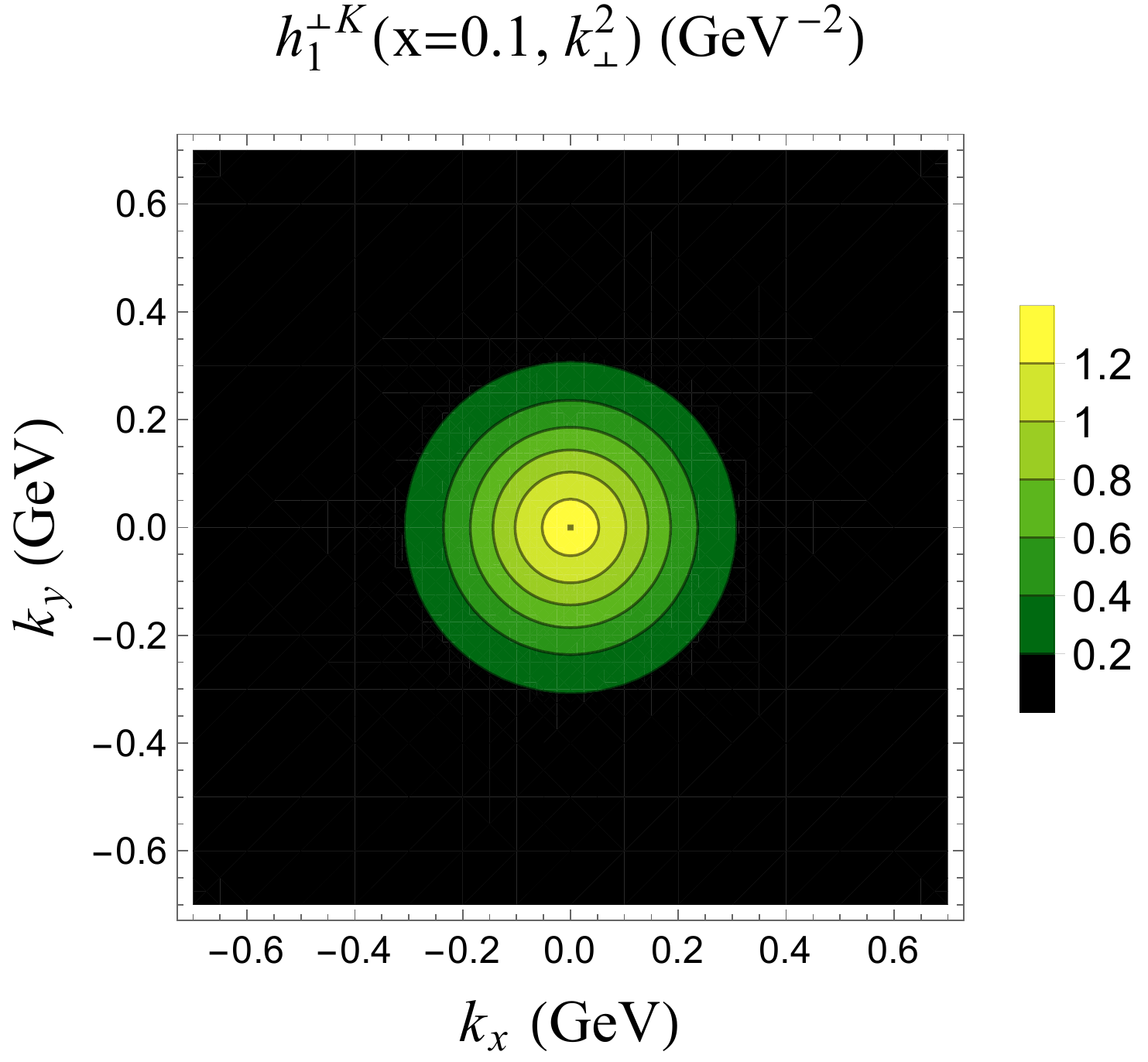}
\hspace{0.05cm}
(e)\includegraphics[width=5.2cm]{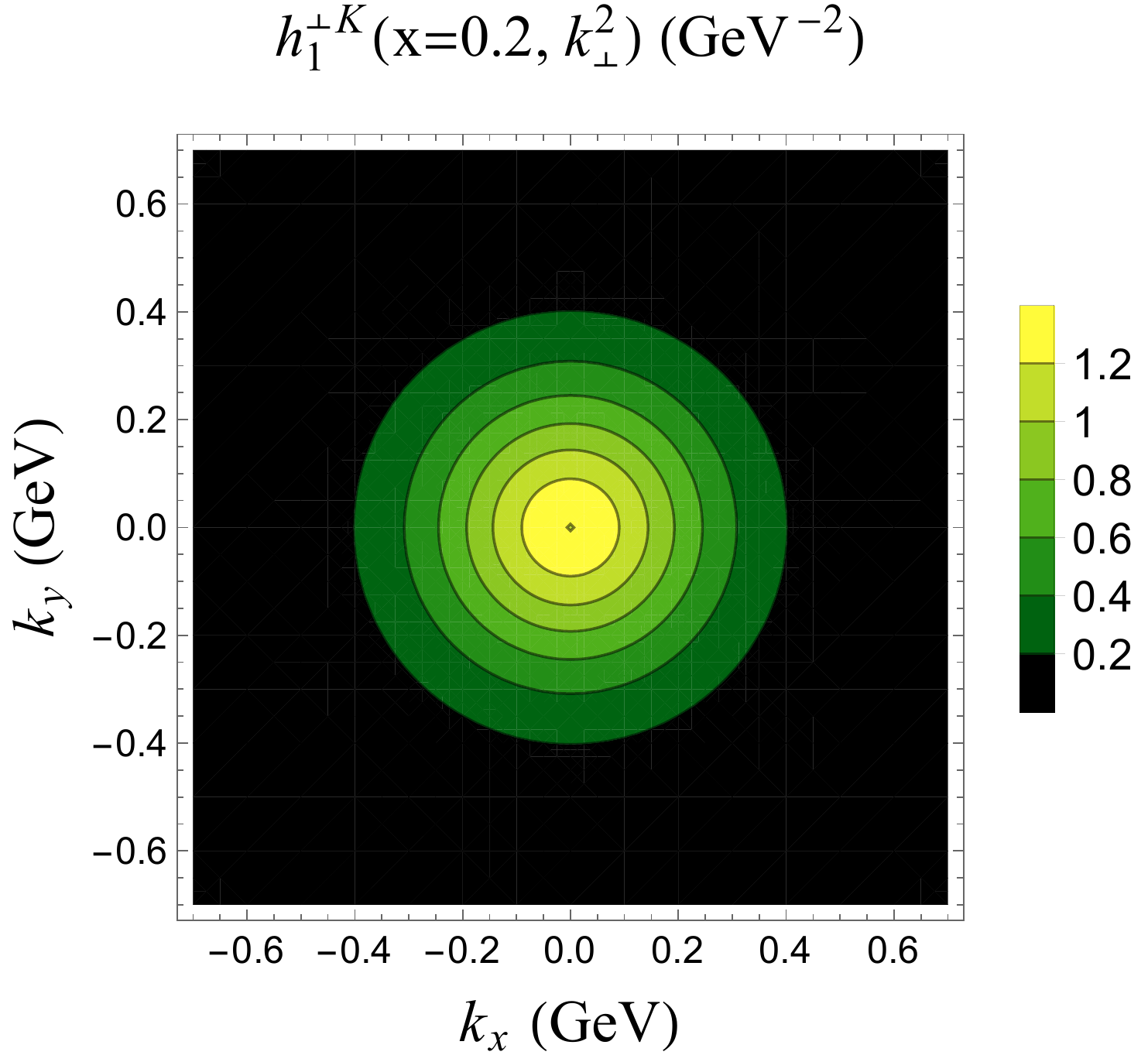}
\hspace{0.05cm}
(f)\includegraphics[width=5.2cm]{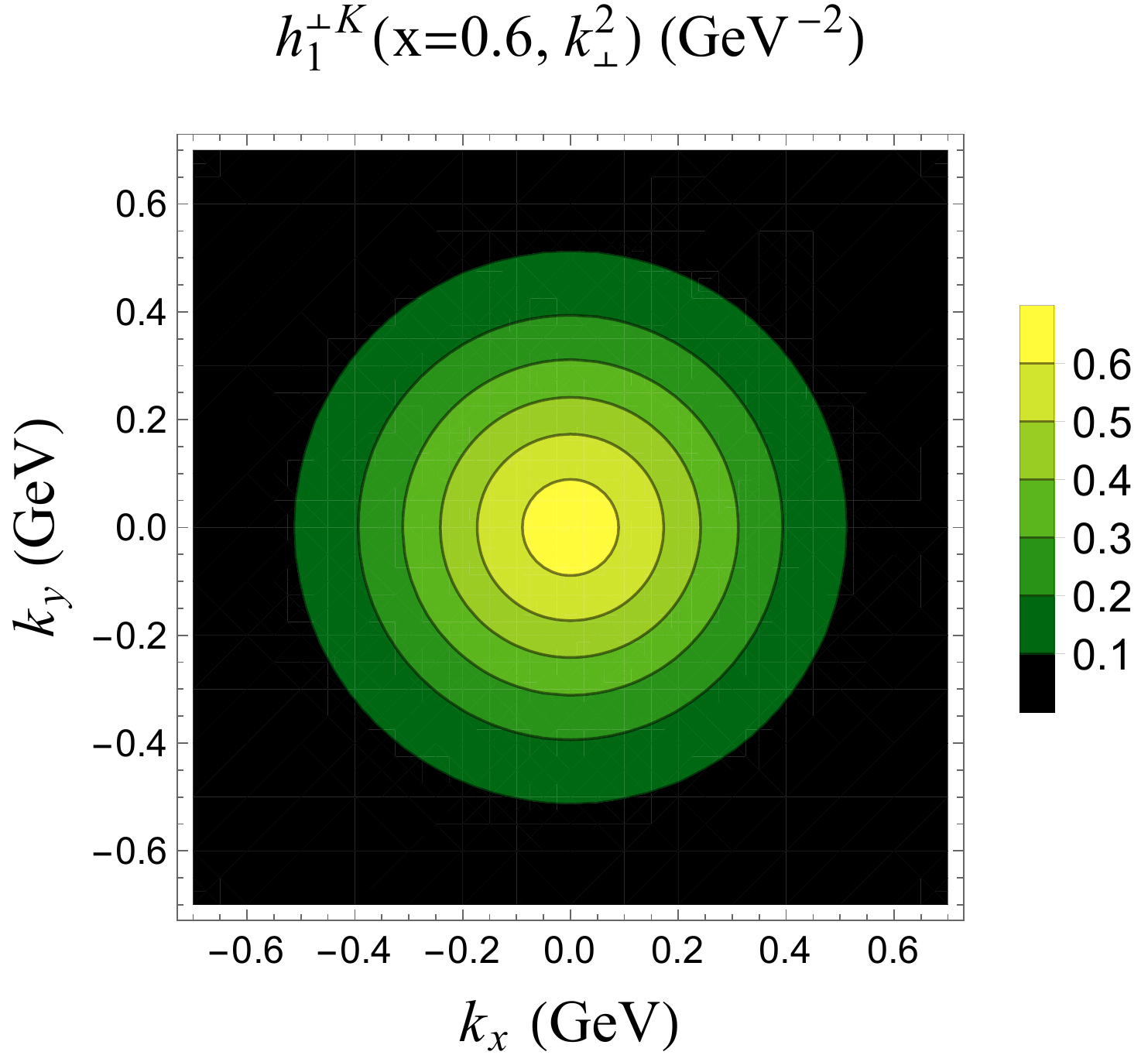}
\hspace{0.05cm}
\end{minipage}
\caption{The Boer-Mulders TMD $ h^{\perp\pi}_1(x,\textbf{k}^2_\perp)$  for pion (upper panel) and kaon (lower panel) at different values of $x$ : $x=0.1$ (left panel), $x=0.2$ (middle panel) and $x=0.6$ (right panel) presented in transverse momentum plane $(k_x, k_y)$.}
\label{bm-pi-tmd}
\end{figure*}

In Fig.~\ref{cp-pi-tmd} and Fig.~\ref{bm-pi-tmd}, we show the unpolarized and the Boer-Mulders TMDs in momentum plane ($k_x,k_y$) by choosing different values of $x$. We observe that the symmetric distribution peak appear to be narrow in transverse momentum plane, while having the lower quark momentum fraction in case of pion and kaon. The unpolarized distributions are found to be wider compared to the polarized distributions.
\subsection{(A) TMD evolution}
The unpolarized TMD evolution is factorized in the framework of Collins–Soper–Sterman (CSS) formalism \cite{evo-collin}. It includes the perturbative affects in the larger energies and momentum transfer regimes, in addition to the non-perturbative affects and the regime of low transverse momentum.  The unpolarized TMD scale evolution is executed in ${\bf b}_\perp$ space which can be done by taking the Fourier transformation of $f_1(x,{\bf k}^2_\perp)$ \cite{tmd-evolution, tmd-evolution1, tmd-evolution2, tmd-evolution3, tmd-evolution4}. We have
\begin{eqnarray}
\tilde{f}_1(x,{\bf b}^2_\perp)= \int_{0}^{\infty} d \textbf{k}_\perp {\bf k}_\perp J_0({\bf k}_\perp {\bf b}_\perp) f_1(x,{\bf k}^2_\perp).
\end{eqnarray}
The TMD evolution of $\tilde{f}_1(x,{\bf b}^2_\perp)$ is given as
\begin{eqnarray}
\tilde{f}_1(x,{\bf b}^2_\perp;\mu)=\tilde{f}_1(x,{\bf b}^2_\perp) R(\mu,\mu_0,{\bf b}_\perp)\ e^{-g_k({\bf b}_\perp) {\rm ln} \frac{\mu}{\mu_0}},
\label{evolution-eqn}
\end{eqnarray}
where the $\mu^2$-evolution operation is directed by the non-perturbative Sudakov factor $g_k({\bf b}_\perp)$ and TMD evolution factor $R(\mu, \mu_0, {\bf b}_\perp)$. We have
\begin{eqnarray}
g_k({\bf b}_\perp)&=& g_2 \frac{{\bf b}^2_\perp}{2},
\end{eqnarray}
where $g_2$ is a free parameter and can be extracted from the experimental data. For the present work, the parameter $g_2$ has been taken from Ref. \cite{bacchetta_evolution} and is given as $g_2=0.13 {\ \rm GeV^2}$. Further, we have
\begin{eqnarray}
R(\mu,\mu_0,{\bf b}_\perp)&=&{\rm exp}\bigg( {\rm ln}\frac{\mu}{\mu_0} \int_{\mu}^{\mu_b} \frac{d \mu'}{\mu'} \gamma_K(\mu') \nonumber\\
&&+ \int_{\mu_0}^{\mu} \frac{d \mu'}{\mu'} \gamma_F\big(\mu',\frac{\mu^2}{{\mu'}^2}\big)\bigg),
\label{parameter-R}
\end{eqnarray}
with $\gamma_K$ and $\gamma_F$ being the anomalous dimensions, which are given as \cite{evo-collin}
\begin{eqnarray}
\gamma_K(\mu')=\alpha_s({\mu'})\frac{c_F}{\pi},
\end{eqnarray}
and
\begin{eqnarray}
\gamma_F\big(\mu',\frac{\mu^2}{{\mu'}^2}\big)=\alpha_s({\mu'})\frac{c_F}{\pi}\bigg(\frac{3}{2}-{\rm ln}\frac{\mu^2}{\mu'^2}\bigg).
\end{eqnarray}
The parameter $\mu_b$ used in Eq. (\ref{parameter-R}) can be expressed in terms of parton impact parameter ${\bf b}_\perp$ as
\begin{eqnarray}
\mu_b=\frac{C_1}{b_*({\bf b}_\perp)},
\end{eqnarray}
where
\begin{eqnarray}
b_*({\bf b}_\perp) =\frac{{\bf b}_\perp}{\sqrt{1+\frac{{\bf b}^2_\perp}{{b^2_{max}}}}} \ \ ;\ \ b_{max}=\frac{C_1}{\mu_0}.
\end{eqnarray}
Here $C_1$ is a constant and we choose its value as $C_1=2 e^{-\gamma_E}$ \cite{evo-collin} with $\gamma_E=0.577$ being the Euler constant.

\begin{figure*}
\centering
\begin{minipage}[c]{1\textwidth}
(a)\includegraphics[width=.42\textwidth]{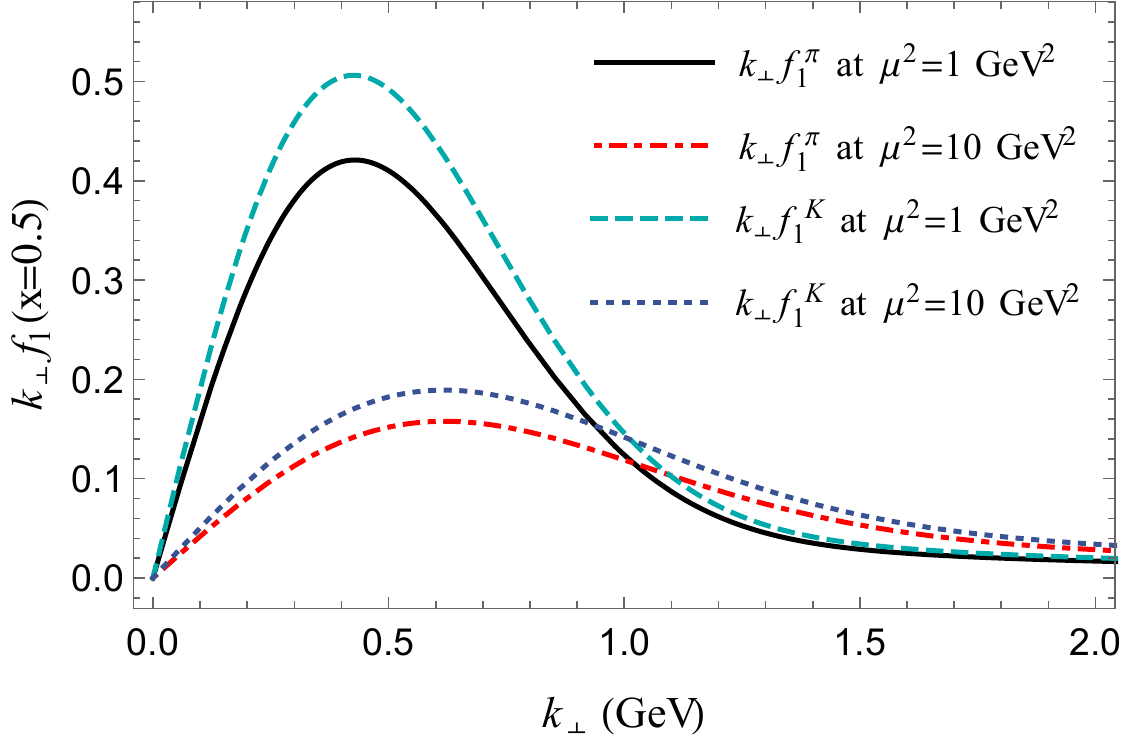}
(b)\includegraphics[width=.42\textwidth]{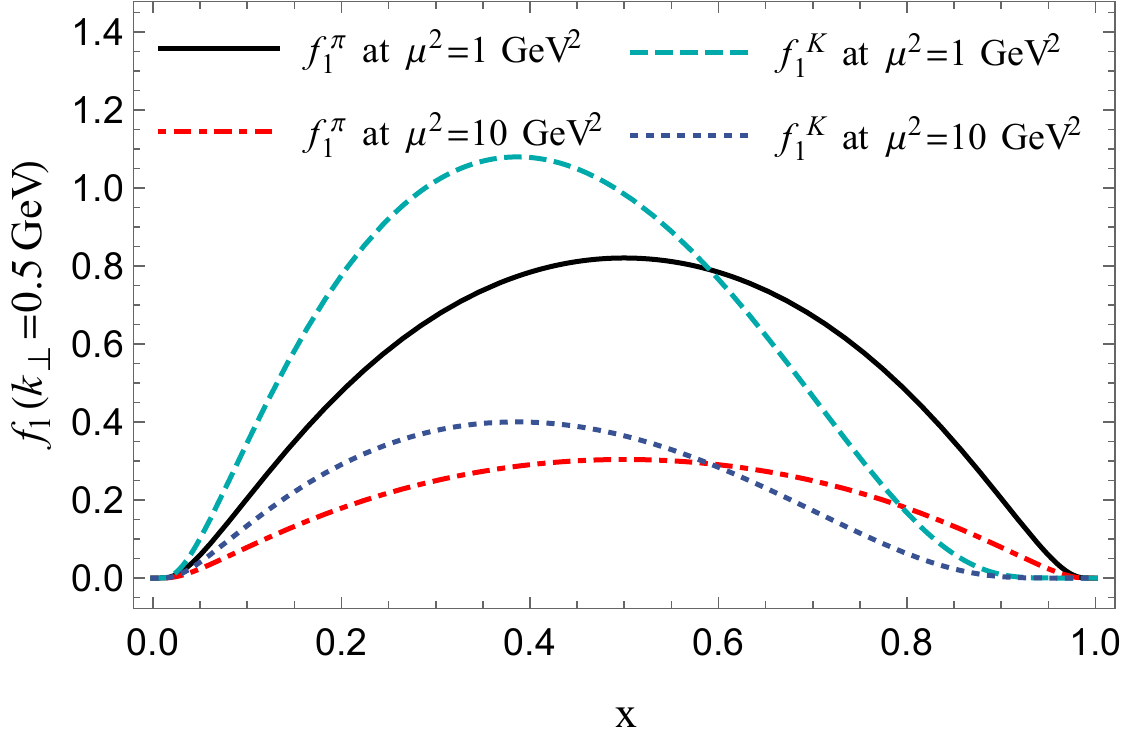}
\end{minipage}
\caption{${\bf k}_\perp f^\pi_1$ and ${\bf k}_\perp f^K_1$ evolved from the initial scale $\mu_0^2=0.246~\rm GeV^2$ to the different scales: $\mu^2=1~{\rm GeV}^2$ and $\mu^2=10~ {\rm GeV}^2$ (a) at fixed $x=0.5$ w.r.t. ${\bf k}_\perp$, (b) at fixed ${\bf k}_\perp$=0.5 GeV w.r.t. $x$.}

\label{tmd-evolution-k}
\end{figure*}

In Fig. \ref{tmd-evolution-k}(a), we provide the graphical representation of unpolarized pion and kaon TMD evolution ${\bf k}_\perp f_1$ w.r.t $x$ by choosing the different scale values $\mu^2=1 {\ \rm and\ } 10 {\ \rm GeV^2}$.
It can be clearly seen that when the TMD evolution is implemented, the distribution peaks become broader and the magnitude goes on decreasing with increase in $\mu^2$.
It would be important to mention here that for the case of pion and kaon, the TMD is evolved from the model scale $\mu^2_0=0.246$ $\rm GeV^2$. 
There is negligible effect of evolution at $\mu=\mu_0$, it simply provide the TMD function multiplied by ${\bf k}_\perp$. With growing $\mu^2$, the width of TMD peaks increases and its values experiences the rapid decrease in magnitude. 
In Fig. \ref{tmd-evolution-k} (b), we observe the broad peak in case of pion as compared to kaon. With the evolution of $f_1$, the magnitude of distribution decreases. Further, we see the clear asymmetry in case of kaon because of heavy spectator antiquark mass.
\section{VII Conclusion} 
We present various quark distributions  in the pion and the kaon in light-cone quark model for the valence quarks suitable for low-resolution properties. 
 The light-cone wave functions in this model have been obtained by transforming the instant-form wave functions through the Melosh-Wigner rotation. We
have obtained reasonable agreement with the experimental data for the pion DA which is also very close to asymptotic DA after LO QCD evolution following ERBL equation. Due to unequal quark masses, we observed distinctly different behavior of kaon DA from the asymptotic DA. The initial scale PDFs have been evaluated using the overlaps of the LCWFs.
We then applied QCD evolution to our
initial pion PDF in order to incorporate degrees of freedom
relevant to higher-resolution probes which allows us to
compare our QCD-evolved PDF with experimental data. The pion PDF at higher scale relevant to E-615 experiment has been computed based on the NNLO DGLAP equations. The initial low-resolution scale is the only adjustable parameter involved in QCD scale evolution. A good agreement with the reanalysis E-615 data has been observed when we evolved the pion PDF from the initial scale $\mu_0^2=0.246$ $\rm GeV^2$.

 Further, we have evaluated GPDs in DGLAP region for zero skewedness i.e. $0<x<1$, which provide us three-dimensional structure of hadron.
 For both the pseudoscalar mesons, depending upon the total momentum transferred to the composite system, we observe the change in distribution with respect to active quark longitudinal momentum fraction. 
 The transverse structure of pion and kaon has also been examined. To evaluate the Boer-Mulders function, we used the perturbative gluon rescattering kernel.
 To observe the combined effect, we have presented the 3D picture of TMD with respect to $x$ and $\textbf{k}^2_\perp$. We observed that as the active quark carries larger longitudinal momentum, the broadening in the transverse momentum plane also increases.  
 Furthermore, we have presented the effect of $\mu^2$ dependence on unpolarized pion and kaon TMDs. We have observed that the magnitudes of the distributions decrease and became wider as $\mu^2$ increases.
 
\section{Acknowledgements} 
H.D. would like to thank the Department of Science and Technology (Ref No. EMR/2017/001549) Government of India for financial support. C.M. is supported by the Natural Science Foundation of China (NSFC) under the Grants No. 11850410436 and 11950410753.


\end{document}